\documentclass[preprint,12pt]{elsart}

\usepackage[english, brazil]{babel}
\usepackage[utf8]{inputenc}
\usepackage{amsmath}
\usepackage{float}
\usepackage{amssymb}
\usepackage{epsfig,psfrag,latexsym,subfigure}
\usepackage{graphics} 
\def\1{{1\kern-.3468em{ 1}}}
\usepackage[active]{srcltx}
\usepackage[color,final]{showkeys}

\usepackage[sort]{natbib}

\newenvironment{proof}[1][Proof]{\begin{trivlist}
\item[\hskip \labelsep {\bfseries #1}]}{\end{trivlist}}

\begin{document}
\selectlanguage{english}
\begin{frontmatter}

\title{\LARGE \bf Modelling and Optimal Control of Multi Strain Epidemics, with Application to COVID-19}

\author[Soton,COPPE]{Edilson F. Arruda\corauthref{cor1}}
\ead{e.f.arruda@southapton.ac.uk}
\author[Cefet]{Dayse H. Pastore},
\ead{dayse.pastore@cefet-rj.br}
\author[FRURJ]{Claudia M. Dias} and 
\ead{mazzaclaudia@gmail.com}
\author[FRURJ]{Shyam S. Das}
\ead{das.shyamsumanta@gmail.com}

\address[Soton]{Department of Decision Analytics and Risk, Southampton Business School, University of Southampton, 12 University Rd, Southampton SO17 1BJ, UK}
\address[COPPE]{Alberto Luiz Coimbra Institute-Graduate School and Research in Engineering, Federal University of Rio de Janeiro. CP 68507, Rio de Janeiro 21941-972, Brasil}
\address[Cefet]{Centro Federal de Educação Tecnológica Celso Suckow da Fonseca, Av. Maracanã, 229, Rio de Janeiro, RJ 20271-110, Brasil}
\corauth[cor1]{Corresponding author. Tel.: +44 023 8059 7677}
\address[FRURJ]{Graduate Program in Mathematical and Computational Modeling, Multidisciplinary Institute, Federal Rural University of Rio de Janeiro, Av. Gov. Roberto Silveira, s/n - Moqueta, Nova Iguaçu  RJ 26020-740, Brasil}

\begin{abstract}
This work introduces a novel epidemiological model that simultaneously considers multiple viral strains, reinfections due to waning immunity response over time and an optimal control formulation. This enables us to derive optimal mitigation strategies over a prescribed time horizon under a more realistic framework that does not imply perennial immunity and a single strain, although these can also be derived as particular cases of our formulation. The model also allows estimation of the number of infections over time in the absence of mitigation strategies under any number of viral strains. We validate our approach in the light of the COVID-19 epidemic and present a number of experiments to shed light on the overall behaviour under one or two strains in the absence of sufficient mitigation measures. We also derive optimal control strategies for distinct mitigation costs and evaluate the effect of these costs on the optimal mitigation measures over a two-year horizon. The results show that relaxations in the mitigation measures cause a rapid increase in the number of cases, which then demand more restrictive measures in the future.
\end{abstract}
\end{frontmatter}

\section{Introduction \label{sec:intro}}

Also known as COVID-19, the Severe Acute Respiratory Syndrome Coronavirus 2 (SARS-CoV-2) is believed to have appeared at the end of 2019 in Wuhan, China \citep{RodriguezClinical2020}. This new, highly transmissible virus spread rapidly around the world, causing significant loss of life and possibly long-lasting economic consequences. The significance of the epidemic prompted a large amount of literature and highlighted the need for comprehensive models combining epidemiology and decision support to help shape public policy; see for example the influential work by \citet{ferguson2020}.

Many mathematical models and data analytic tools have been proposed to understand the evolution of the COVID-19 pandemic throughout the world, generally based on classical epidemiological models \citep{Kermack1927,Kermack}. For example, \citet{ferguson2020} promoted non-pharmaceutical interventions and \citet{flaxman2020} evaluated the effect of such measures in Europe. Later, \citet{Tarrataca2021} evaluated the effectiveness of long-term on-off lock-down policies, whilst \citet{Kantner2020} pursued optimal trade-offs between economics and healthcare concerns. Like most of the literature, these works did not consider the possibility of reinfection or multiple viral strains. Similarly, these possibilities were also disregarded in investigations of optimal strategies to exit lock-down which also did not consider the possibility of multiple waves of infection \citep{Ruktanonchaieabc2020,Rawson2020}.

Also essential to shape public policy and prevention and treatment strategies is a thorough understanding of the mechanism of the pandemic. This includes the genomics mapping of viral strains \citep{Callaway2020,KORBER2020812}, which when carried out in Brazil revealed more than 100 viral strains of COVID-19 \citep{resende2020, vieira2020,Voloch2020}, three of which managed to survive. Such a reduction in genetic diversity has been attributed to the social isolation measures in that country \citep{anatereza}. Distinct strains have also been recently identified in the United Kingdom \citep{KIRBY2021} and South Africa \citep{Tegally2020} which have rapidly spread around the globe. Further studies are needed to properly assess the mortality rate of these new variants, but the UK strain is already believed to be around 60 to 70\% more transmissible than the original variant.

Another important challenge to modellers is that the immune response to COVID-19 is not uniform \citep{long_2020}, may reportedly wane over time \citep{Seow2020,Daneabf4063,Edridge_2020} and reinfection is possible \citep{BONIFACIO2020,Tillett_2020}. Furthermore, the same patient may be infected by different strains of the virus \citep{Nonaka2021,to_2020}. \citet{Overbaugh2020} argues that a thorough understanding of reinfection is essential for understanding the spread of the disease, whereas \citet{DAWOOD2020100673} foretells future global challenges to contain epidemics with reinfection. From a more operational standpoint, a recent work made use of available databases and the classical SIR (susceptible, infected, recovered) framework to estimate the number of COVID-19 reinfections from empirical data \citep{McMahon2020}.

Although COVID-19 reinfection and multiple viral strains have received increased attention in the literature, mathematical modelling that incorporates these characteristics is still scarce. \citet{Khyar2020} searched for stability conditions within a general two-strain model and assessed the effect of a quarantine strategy to curb COVID-19 spread in Morocco. More generally, viral reinfection is often studied with emphasis on stability conditions and disease-free equilibrium \citep[e.g.,][]{Perthame2003}. In particular, \citet{Etbaigha2018} proposed a SEIR (susceptible, exposed, infected, removed) model for swine influenza and analysed the effect of prescribed vaccination strategies. Finally, a simpler SIR model is employed in \citep{Fudolig2020} to study the dynamics of two viral strains considering that the second strain appears after the first strain reaches equilibrium. In general, whilst these models examine long-term stability, they do not incorporate decision support tools and optimisation.

To support decision making, optimal control approaches have been proposed to promote compromises between COVID-19 infection levels and economic consequences of non-pharmaceutical interventions \citep{Kantner2020,Bursac2020,Perkins2020}. The control mechanism may consist of a proportional reduction in the infection levels \citep{Kantner2020,Perkins2020} or include quarantine, isolation and public health education \citep{Bursac2020}. Even though these models do not account for reinfection and multiple viral strains, they do provide interesting insights. Perhaps the most interesting insight is that high levels of control are needed from the beginning to preserve healthcare systems and leverage control options late in the epidemics \citep{Perkins2020}. This is consistent with the empirical results in \citep{Tarrataca2021}.

Whereas models considering multiple viral strains are rare, one can still find in the literature optimal control approaches based on classical epidemiological models for two viral strains \citep{Bentaleb2020,Gubar2017}. These are general epidemiological models, i.e. not specifically tailored for a given epidemic, that do not consider reinfection. A limiting feature of the model of \citet{Bentaleb2020}, however, is that it relies on a curative treatment. In contrast, the discrete network-based model of \citet{Gubar2017} relies on individual control measures to be applied separately to each of the two strains.

To the best of our knowledge, this is the first paper to simultaneously consider multiple viral strains, reinfection and optimal control. Amongst the novel contributions of this work, we generalise the  preceding literature \citep{Bentaleb2020,Gubar2017,Khyar2020} by considering not only two but any number of viral strains. Based on the SEIR framework, the model innovates by accounting for the loss of immunity over time and contemplating the possibility of reinfection, which has the potential to considerably increase the infection levels over time. Finally, we propose a novel optimal control approach whereby a proportional reduction of the infection rate by mitigation measures (such as non-pharmaceutical interventions) incurs an exponentially increasing cost. We argue that this approach is more realistic than assuming linear or quadratic costs \citep[e.g.,][]{Bursac2020,Perkins2020}, once it has become clear that reduction in transmission is increasingly difficult to obtain, and therefore increasingly more costly, once mitigating measures are already in place. The proposed approach seeks for a compromise between the overall number of deaths and the intervention costs over a prescribed horizon.

In addition to the methodological innovations, we also contribute by providing a more realistic framework for epidemic modelling that avoids the sometimes optimistic assumptions of perennial immunity and a single viral strain. The framework also includes an optimal control formulation that enables the decision makers to clearly define the compromises between loss of life and economic consequences over a prolonged time horizon.

The remainder of this paper is organised as follows. Section \ref{sec:model} introduces the multi-strain model with reinfection and Section \ref{sec:equi} analyses its equilibrium points and the reproductive number. Section \ref{optimal} proposes a novel optimal control formulation for the multi-strain model, which is solved to derive the optimal control strategy over a prescribed time horizon. Section \ref{sec:experimental} features a series of experiments designed to illustrate the system's behaviour in the presence of one and two strains and with insufficient mitigation measures. Furthermore, it also derives and interprets optimal control strategies over a two-year horizon under distinct mitigation costs. Finally, Section \ref{sec:conclusion} concludes the paper.

\section{Proposed Mathematical Model \label{sec:model}}


Let $V = \{1, \, \ldots, \, n\}$ be the set of virus strains circulating in the population, and let $j \in V$ denote a particular strain. For each $j \in V$ and time $t \ge 0$, let $S_j(t), \, E_j(t), \, I_j(t)$ and $R_j(t)$, respectively denote the number of susceptible, exposed, infected and removed (recovered and immune) individuals in the population at time $t$. In addition, $P(t)$ denotes the total population at time $t \ge 0$.  

The susceptible population $S_j(t)$ includes all individuals that are not immune strain $j \in V$ at time $t \ge 0$ and therefore can catch the disease. In turn, $E_j(t)$ comprises all individuals that have been recently contaminated by strain $j$ but are currently in the latency period and therefore have not yet manifested the disease and become infectious. Finally, $I_j(t)$ counts all individuals that have caught and manifested the strain $j$ and are still suffering from it, whereas $R_j(t)$ denotes the total number of individuals that are recovered and immune to strain $j$ at time $t$.

\begin{figure}[htbp!]
	\centering
	\includegraphics[scale=0.5]{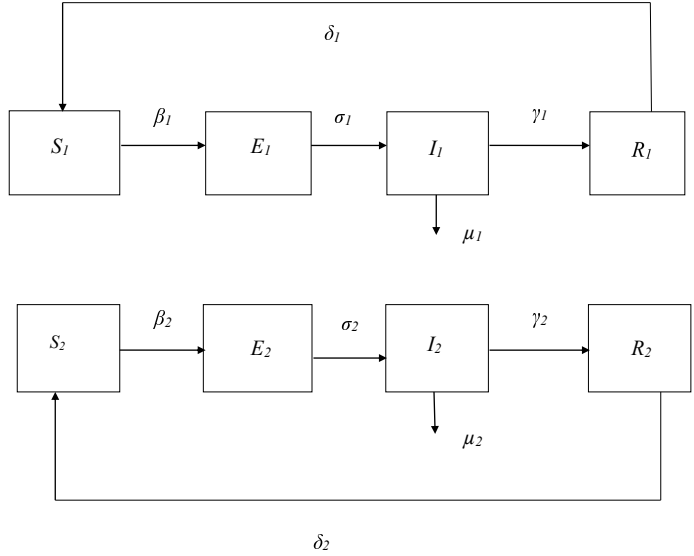}
	\caption{Schematic diagram of the proposed model for two virus strains. \label{fig:diagram} }
\end{figure}

The proposed multi strain model follows Eq. \eqref{Eq1-1}-\eqref{Eq4} below: 


\begin{align}
\dot P(t) &= -\sum_{j=1}^n \mu_j I_j \label{Eq1-1}\\
S_j(t) &= P(t) - E_j(t) - I_j(t) - R_j(t) \label{Eq4-1}\\
\dot{E_{j}}(t) &= (\, 1-u(t) \,)\beta_{j} S_{j}(t) I_{j}(t) - \sigma_{j}E_{j}(t), \label{Eq2}\\
\dot{I_{j}}(t) &= \sigma_{j} E_{j} (t) -(\mu_{j}+\gamma_j)I_{j}(t), \label{Eq3} \\
\dot{R_{j}}(t) &= \gamma_jI_{j}(t) - \delta_j R_j(t), \label{Eq4}
\end{align}
where $S_j(0) \le P(0), \, \forall \; j \in V$. For the sake of illustration, Figure \ref{fig:diagram} represents the dynamics of a two-strain model.

Consider the dynamics of a given strain $j \in V$. Observe from Eq. \eqref{Eq2} that susceptible individuals can contract this strain when in contact with a contagious carrier belonging to the infected population. The rate of infection is $\beta_j > 0$ and $u(t) \in [0,1]$ emulates the lock-down effect at time $t \ge 0$: $u(t) = 1$ indicates 100\% effective mitigating measures and $u(t) = 0$ represents the absence of non-pharmaceutical interventions, whereas $u(t) \in (0,1)$ indicates partially effective measures to limit the spread of the disease. The first term in the right hand side of \eqref{Eq2} represents the formerly susceptible individuals that have just been infected, whereas the second term indicates the exposed individuals that have just manifested the once latent disease. The latter enter the infected compartment in the right hand side of Eq. \eqref{Eq3}. The second term in the right hand side of \eqref{Eq3} represents infected individuals that recover - at rate $\gamma_j > 0$, or die - at rate $\mu_j \ge 0$. Finally, each newly recovered individual moves to the \emph{removed} compartment - first term of the right hand side of \eqref{Eq4}. The second term in the right hand side of \eqref{Eq4} represents the loss of immunity over time. 
Finally, Eq. \eqref{Eq4-1} keeps track of the individuals that are currently susceptible to strain $j \in V$, whereas Eq. \eqref{Eq1-1} monitors the evolution of the total population over time. Table \ref{tab:parameters} describes the system's parameters.

\begin{table}[htb]
\caption{Parameters for multi-strain dynamics. \label{tab:parameters}}
\centering
\small{
\begin{tabular}{|c|l|c|} \hline \hline
\textbf{Parameter} & \textbf{Description} & \textbf{Unit} \\ \hline
$\beta_j$ & Transmission rate of strain $j$ & transmissions/encounter \\ \hline
$\sigma_j$ & Inverse of the latency period of strain $j$ & $\text{days}^{-1}$ \\ \hline
$\gamma_j$ & Recovery rate for strain $j$ & $\text{days}^{-1}$ \\ \hline
$\delta_j$ & Rate of immunity loss for strain $j$ & $\text{days}^{-1}$ \\ \hline
$\mu_j$ & Death rate due to strain $j$ & $\text{days}^{-1}$ \\ \hline
$u(t)$ & Mitigation (lock-down) effect at time $t$ & - \\ \hline
\end{tabular}
}
\end{table}

\section{The Equilibrium Points \label{sec:equi}}

To simplify our analysis, in this section we assume a constant control, i.e. $u(t) = u \in [0,1], \, \forall t \ge 0$.  A simple inspection to the system of equations \eqref{Eq1-1}-\eqref{Eq4} yields
\begin{equation}\label{eq:suscetivel}
\dot{S_{j}}(t) = -(\, 1-u \,)\beta_{j}S_{j}(t) I_{j}(t) + \delta_j R_j(t)- \sum^{n}_{i=1; i\neq j} \mu_i I_i(t).
\end{equation}

Hence, it is not hard to verify that the trivial equilibrium point is the infection free point, with 
\begin{equation} \label{eq:trivial}
    {E}_{j}(\infty) = {I}_{j}(\infty)= {R}_{j}(\infty)=0, \quad S_j(\infty) = \overline{S}_j \ge 0, \quad P(\infty) \ge 0. 
\end{equation}

To calculate the non-trivial equilibrium, we start with the case of two strains below.

\begin{thm} \label{theo:non-trivial}
Suppose $|V| = 2$. Then, besides the trivial equilibrium point in Eq. \eqref{eq:trivial}, the system has a non-trivial equilibrium point with $I_1 \le 0$.
\end{thm}
\begin{proof}
Since $|V| = 2$, we have exactly two roots, one of which is the trivial equilibrium. For the non-trivial equilibrium, we must have $I_1 \neq 0$ and $I_2 \neq 0$. Hence, by equalling the left-hand side of \eqref{Eq1-1}-\eqref{Eq4} to zero, we obtain:
\begin{align*}
S_1(\infty) =  \overline{S}_1 = \frac{\mu_1+\gamma_1}{(\, 1-u \,)\beta_1}, \quad 
S_1(\infty) = \overline{S}_2 = \frac{\mu_2+\gamma_2}{(\, 1-u \,)\beta_2} \\
E_1(\infty) = \overline{E}_1 =-\frac{(\mu_1+\gamma_1)\mu_2 \overline{I}_2}{\mu_1 \sigma_1}, \quad
E_2(\infty) = \overline{E}_2 = \frac{(\mu_2+\gamma_2) \overline{I}_2}{\sigma_2}, \\ 
I_1(\infty) = \overline{I}_1 =-\frac{\mu_2 \overline{I}_2}{\mu_1}, \quad I_2(\infty) = \overline{I}_2, \\
R_1(\infty) = \overline{R}_1 = -\frac{\gamma_1 \mu_2 \overline{I}_2}{\mu_1 \delta_1}, \quad R_2(\infty) = \overline{R}_2 =\frac{\gamma_2\overline{I}_2}{\delta_2}
\end{align*}
Since all coefficients are positive, it follows that $\overline{I}_1 \le 0$.
\end{proof}

From Theorem \ref{theo:non-trivial}, it follows that the non-trivial equilibrium point of a two-strain model is biologically infeasible, and therefore of no practical interest. Theorem \ref{theo:non-trivial-multi} below generalises this result for multiple strains, i.e. $|V| > 2$.

\begin{thm} \label{theo:non-trivial-multi}
Suppose $|V| = n >  2$. Then, besides the trivial equilibrium point in Eq. \eqref{eq:trivial}, the system has a non-trivial equilibrium point with $I_j \le 0, \forall j \in \{1, \, \ldots, \, n\}$.
\end{thm}
\begin{proof}
Making the left hand side of Eq. \eqref{Eq3} equal to zero yields $ E_{j} =\frac{ \mu_{j}+\gamma_j}{\sigma_{j}}I_{j}$. Replacing this result in Eq. \eqref{Eq2}, we obtain:
\begin{gather*}
   0 =(\, 1-u \,)\beta_{j} S_{j} I_{j} - \sigma_{j}E_{j}  =(\, 1-u \,)\beta_{j} S_{j} I_{j} -({ \mu_{j}+\gamma_j})I_{j} \\
   0 = I_j \left( (\, 1-u \,)\beta_{j} S_{j} -({ \mu_{j}+\gamma_j}) \right).
\end{gather*}
Therefore, either $I_j = 0$ or $S_j = \frac{\mu_j+\gamma_j}{(\, 1-u \,) \beta_j}$. Now, substituting the latter equality in \eqref{eq:suscetivel} and making the derivative nil, we have:

%
\begin{align*}
0 & = -(\, 1-u \,)\beta_{j}S_{j} I_{j} + \delta_j R_j- \sum^{n}_{i=1; i\neq j} \mu_i I_i \\
& =  -(\, 1-u \,)\beta_{j}\frac{\mu_j+\gamma_j}{(\, 1-u \,) \beta_j} I_{j} + \delta_j R_j- \sum^{n}_{i=1; i\neq j} \mu_i I_i\\
& \implies 0= -(\mu_j+\gamma_j) I_{j} + \delta_j R_j- \sum^{n}_{i=1; i\neq j} \mu_i I_i \\
& \implies 0=-{\gamma_j} I_{j} + \delta_j R_j- \sum^{n}_{i=1} \mu_i I_i\\
& \text{ eq. (\ref{Eq4})} \implies 0= \sum^{n}_{i=1} \mu_i I_i.
\end{align*}
Since $\mu_j>0 \; \forall j$, for the last equality to hold we must have either $I_j = 0, \, j= 0, \, 1\ \, \ldots n$, or $I_j < 0$ for at least one $j \in \{1, \, \ldots, \, n$\}.
\end{proof}

Theorem \ref{theo:non-trivial-multi} therefore implies that the non-trivial equilibrium point is biologically infeasible and of no practical use for any number of different strains. In the remainder of this paper, we will only consider biologically feasible solutions.


\subsection{Stability \label{sec:stability}}

Considering that only the trivial equilibrium points are of biological interest, this section analyses the stability solely with respect to these points. The system \eqref{Eq1-1}-\eqref{Eq4} has a dimension $4\times n +1$. Consequently, the Jacobian matrix associated with the system and applied to the trivial equilibrium is of order $(4\times n +1)^2$. It has $n+1$ null eigenvalues and $n$ eigenvalues equal to $-\delta_j, j=1,...,n$. The remaining $2 \times n$ eigenvalues are given by:
\begin{align*}
-1/2(\mu_j+\gamma_j+\sigma_j)+1/2\,\sqrt {4\,(\, 1-u \,)\,\beta_j\,
\sigma_j \,\overline{S}_j + \left( \mu_j+\gamma_j-\sigma_j\right) ^{2}},\\
-1/2(\mu_j+\gamma_j+\sigma_j)-1/2\,\sqrt {4\,(\, 1-u \,)\,\beta_j\,
\sigma_j \,\overline{S}_j + \left( \mu_j+\gamma_j-\sigma_j\right) ^{2}}.
\end{align*}

To prove stability we need to show that the real part of the eigenvalues are negative. Therefore, it suffices to show that
\[
-1/2(\mu_j+\gamma_j+\sigma_j)+1/2\,\sqrt {4\,(\, 1-u \,)\,\beta_j\,
\sigma_j \,\overline{S}_j + \left( \mu_j+\gamma_j-\sigma_j\right) ^{2}}<0,
\]
since this implies that the remaining eigenvalues will also be negative. The expression above holds if:
\[
\,\sqrt {4\,(\, 1-u \,)\,\beta_j\,
\sigma_j \,\overline{S}_j + \left( \mu_j+\gamma_j-\sigma_j\right) ^{2}}<(\mu_j+\gamma_j+\sigma_j),
\]
\[
{4\,(\, 1-u \,)\,\beta_j\,
\sigma_j \,\overline{S}_j + \left( \mu_j+\gamma_j-\sigma_j\right) ^{2}}<(\mu_j+\gamma_j+\sigma_j)^2,
\]
\[
{4\,(\, 1-u \,)\,\beta_j\,
\sigma_j \,\overline{S}_j + \left( \mu_j+\gamma_j-\sigma_j\right) ^{2}}<(\mu_j+\gamma_j+\sigma_j)^2,
\]
\[
(\, 1-u \,)\,\beta_j\,\overline{S}_j <(\mu_j+\gamma_j).
\]

From the latter inequality, we can define the reproduction number,
\begin{equation}\label{eq:R0}
R_0=\max_{j=1,...,n}{\frac{(\, 1-u \,)\,\beta_j\,\overline{S}_j}{\mu_j+\gamma_j}}.
\end{equation}

We can say that the trivial equilibrium point (without infection) is locally asymptotically stable if $R_0<1$. Hence, Eq. \eqref{eq:R0} implies a minimum level of constant lock-down effect $u \in [0,1]$ to stabilise the system. Observe that, since the lock-down effect applies to all viral strains, it suffices to stabilise the system with respect to the most transmissible strain. 

In the next section, we expand the analysis to search for time varying lock-down effects with a view to optimising the long-term cost of non-pharmaceutical (lock-down) interventions.

\section{Optimal Lock-down Strategies \label{optimal}}

To control the spread of the disease in the population, the proposed strategy considers an isolation level of the population $u(t), \, t \ge 0$ at any time $t$. To account for the time-varying control, let us rewrite the system of equations \eqref{Eq1-1}-\eqref{eq:suscetivel} as follows:
\begin{align}
\dot P(t) &= -\sum_{j=1}^n \mu_j I_j \label{Eq5}\\
\dot{S_{j}}(t) &= -(1-u(t))\beta_{j}S_{j}(t) I_{j}(t) + \delta_j R_j(t)- \sum^{n}_{i=1; i\neq j} \mu_i I_i \label{eq:susc}\\
\dot{E_{j}}(t) &= (1-u(t))\beta_{j} S_{j}(t) I_{j}(t) - \sigma_{j}E_{j}(t), \label{Eq7}\\
\dot{I_{j}}(t)& = \sigma_{j} E_{j} (t) -(\mu_{j}+\gamma_j)I_{j}(t), \label{Eq8} \\
\dot{R_{j}}(t) &= \gamma_jI_{j}(t) - \delta_j R_j(t), 
 \label{Eq6}
\end{align}

To find a meaningful trade-off between the cost of the control, i.e. lock-down measures or non-pharmaceutical interventions, and the cost of elevated infection levels to the healthcare system and the population in general, we define the following functional cost:
\begin{equation} \label{eq:functional}
   J(P,u) = c_1 P -  e^{c_2 u}, \; 0 \le u \le 1,
\end{equation}
where $c_1 > 0$ and $c_2 > 0$ are scalar parameters.\\

Recall that in the revised formulation Eq. \eqref{Eq5}-\eqref{Eq6},  $u(t) = 0$ indicates no lock-down and $u(t) = 1$ corresponds to full lock-down. Observe that the cost in \eqref{eq:functional} grows with the population size and decreases as a function of the control $u$. While increasing $u$ decreases the functional, it also implies a decrease in the number of infections and, therefore, deaths. And less deaths imply in an increased total population, thus increasing the functional. Observe also that the cost of control increases exponentially in the feasible interval $[0,1]$, to mimic the fact that extra mitigation measures tend to become increasingly costly.

Let $\psi = \{u(t), \, t \in(0,T): u(t) \in [0,1]$\} be a feasible lockdown strategy and let $\Psi$ denote the set of all feasible strategies. For each control strategy $\psi \in \Psi$, let  
\begin{equation} \label{eq:functional-integral}
   J(\psi) = \int_0^T J( P(s), u(s)) \,ds
\end{equation}
denote the overall cost of the strategy. The optimal control problem then becomes:
\begin{equation} \label{eq:optimal-control}
\begin{array}{l}
   \text{Maximise} \;J(\psi), \, \psi \in \Psi \\
   \text{subject to} \; \eqref{Eq5}-\eqref{Eq6}.
   \end{array}
\end{equation}

The overall objective in \eqref{eq:optimal-control} is to minimise the number of deaths over time, which is equivalent to maximising the population, whilst also accounting for the cost of lock-down measures represented by the negative term in \eqref{eq:functional}.


\subsection{Solution of the Optimal Control Problem \label{optimal-sol}}

The solve \eqref{eq:optimal-control}, we make use of Pontryagin's maximum principe \citep{Kirk1970,Bryson1975,Pontryagin1962}. Firstly, we need to formulate the Hamiltonian function of our optimal control problem, given by:
%
\begin{multline}
  H =  c_1 P -  e^{c_2 u} + \phi_{P}\dot{P} + \\ 
  +\sum_{j=1}^{n} \phi_{S_j}\dot{S}_j +\sum_{j=1}^{n} \phi_{E_j}\dot{E}_j +\sum_{j=1}^{n} \phi_{I_j}\dot{I}_j+\sum_{j=1}^{n} \phi_{R_j}\dot{R}_j + \eta u.
\end{multline}

In the above equation, $\phi_{P}$ represents the co-state variable corresponding to the original variable $P$; similarly, the subscript of the remaining co-state variables $\phi_{ \_ }$ indicates the corresponding original variable. In addition, $\eta \ge 0$ is a penalty multiplier added to ensure that $u \ge 0$; at optimality we must have $\eta u^* = 0$. By deriving the Hamiltonian with respect to the original variables in \eqref{Eq5}-\eqref{Eq6}, we obtain adjoint system of equations with respect to the co-state variables:
\begin{align*}
\frac{d \phi_P}{dt} =\, - \frac{\partial H}{\partial P}=\,& -c_1 \\
\frac{d\phi_{S_j}}{dt}=\, - \frac{\partial H}{\partial S_j}=\,&   ( \phi_{S_j} - \phi_{E_j} ) \,(1-u)\,\beta_j I_j \\
\frac{d \phi_{E_j}}{dt}=\, - \frac{\partial H}{\partial E_j}=\,& \sigma_j (\phi_{E_j} -\phi_{I_j}) \\ 
\frac{d \phi_{I_j}}{dt}=\, - \frac{\partial H}{\partial I_j}=\,& (\phi_{S_j} -\phi_{E_j}) (1-u) \beta_j S_j +\phi_{I_j} (\mu_j+\gamma_j)-\phi_{R_j} \gamma_j+\phi_P \mu_j +\\
&+ \mu_j(\sum^{n}_{i=1; i\neq j} \phi_{S_i}) \\
\frac{d \phi_{R_j}}{dt}=\, - \frac{\partial H}{\partial R_j}=\,&\delta_j (\phi_{R_j}-\phi_{S_j})
\end{align*}
with transversality conditions,
\[
\phi_{P}(T)=\phi_{S_j}(T)=\phi_{E_j}(T)= \phi_{I_j}(T)= \phi_{R_j}(T)=0,\, \, \forall \, j=1, \cdot \cdot \cdot, n.
\]

\begin{thm} \label{thm:optimalcontrol}
The solution to the optimal control problem in \eqref{eq:optimal-control} yields:
\begin{equation} \label{eq:optimalcontrol}
u^*=\max \left\{0,\frac{1}{c_2}\ln\left(\frac{1}{c_2} \displaystyle \sum_{j=1}^{n} S_j I_j \beta_j (\phi_{S_j}-\phi_{E_j})\right)  \right\}.
\end{equation}
\end{thm}
\begin{proof}
The optimal solution $u^*$ must satisfy:
\[
 \frac{\partial H}{\partial u^*}=\, -c_2e^{c_2u^*}+ \sum_{j=1}^{n} S_j I_j \beta_j (\phi_{S_j}-\phi_{E_j}) + \eta =0.
\]
Thus, isolating $u^*$, we obtain,
\[
c_2e^{c_2u^*}= \sum_{j=1}^{n} S_j I_j \beta_j (\phi_{S_j}-\phi_{E_j}) + \eta 
\]
\[
e^{c_2u^*}= \frac{\sum_{j=1}^{n} S_j I_j \beta_j (\phi_{S_j}-\phi_{E_j}) + \eta }{c_2}
\]
\[
c_2u^*=\ln{\left( \frac{\sum_{j=1}^{n} S_j I_j \beta_j (\phi_{S_j}-\phi_{E_j}) + \eta }{c_2}\right)}
\]
\[
u^*=\frac{1}{c_2} \ln{\left( \frac{\sum_{j=1}^{n} S_j I_j \beta_j (\phi_{S_j}-\phi_{E_j}) + \eta }{c_2}\right)}
\]
If $u^* >0$, we necessarily have $ \eta = 0$, since  $\eta u^* = 0$. Consequently, the optimal control can be expressed as:
\begin{equation} \label{eq:cont}
u^*=\frac{1}{c_2} \ln{\left( \frac{\sum_{j=1}^{n} S_j I_j \beta_j (\phi_{S_j}-\phi_{E_j})}{c_2}\right)}.
\end{equation}

But, if $u^*=0$, then
\[
u^*=\frac{1}{c_2} \ln{\left( \frac{\sum_{j=1}^{n} S_j I_j \beta_j (\phi_{S_j}-\phi_{E_j}) + \eta }{c_2}\right)}=0,
\]
\[
\ln{\left( \frac{\sum_{j=1}^{n} S_j I_j \beta_j (\phi_{S_j}-\phi_{E_j}) + \eta }{c_2}\right)}=0,
\]
which, considering the properties of the logarithm function, yields
\[
 \frac{\sum_{j=1}^{n} S_j I_j \beta_j (\phi_{S_j}-\phi_{E_j}) + \eta }{c_2}=1,
\]
\[
 \sum_{j=1}^{n} S_j I_j \beta_j (\phi_{S_j}-\phi_{E_j}) + \eta=c_2,
\]
\[
 \eta=c_2- \sum_{j=1}^{n} S_j I_j \beta_j (\phi_{S_j}-\phi_{E_j}) > 0.
\]
The inequality above holds true because, by definition $\eta u* = 0$ and $\eta \ge 0$; the case where $\eta = 0$ was already explored in Eq. \eqref{eq:cont}.
Hence, the expression below summarises the optimal control results:
\[
u^*=\max \left\{0,\frac{1}{c_2}\ln\left(\frac{1}{c_2} \displaystyle \sum_{j=1}^{n} S_j I_j \beta_j (\phi_{S_j}-\phi_{E_j})\right)  \right\}.
\]
\end{proof}

\section{Numerical Experiments \label{sec:experimental}}

In order to better understand the long-term behaviour of the system \eqref{Eq1-1}-\eqref{Eq4}, we performed a simple experiment with a single virus strain, which we will call \emph{Experiment 1}. The parameters for this experiment are based on \citep{Tarrataca2021} and appear in Table \ref{tab:parex1} below. Note that  \emph{Experiment 1} does not consider any lock-down effect, which means that $u(t) = 0, \forall t \ge 0$.

\begin{table}[htb]
\caption{Parameters for \emph{Experiment 1.} \label{tab:parex1}}
\scriptsize{
\centering
\begin{tabular}{|c|c|} \hline \hline
\textbf{Parameter} & \textbf{Value} \\ \hline
$\beta_1$ &  $2.41 \cdot 10^{-9}$ \\ \hline
$\sigma_1$ & $\frac{1}{7} \, \text{days}^{-1}$ \\ \hline
$\gamma_1$ & $\frac{1}{21} \, \text{days}^{-1}$ \\ \hline
$\delta_1$ & $\frac{1}{90} \, \text{days}^{-1}$ \\ \hline
$\mu_1$ & $1.152 \cdot 10^{-5} \,\text{days}^{-1}$ \\ \hline
$u(t)$ & 1.0 \\ \hline
\end{tabular}
\hspace{1cm}
\begin{tabular}{|c|} \hline \hline
\textbf{Initial Conditions} \\ \hline
$S(0) = 217 \cdot 10^6$ \\ \hline
$E(0) = 252$ \\ \hline
$I(0) = 2$ \\ \hline
$R(0) = 1$ \\ \hline
\end{tabular}
}
\end{table}

\begin{figure}[htbp!]
	\centering
	\includegraphics[width=\linewidth]{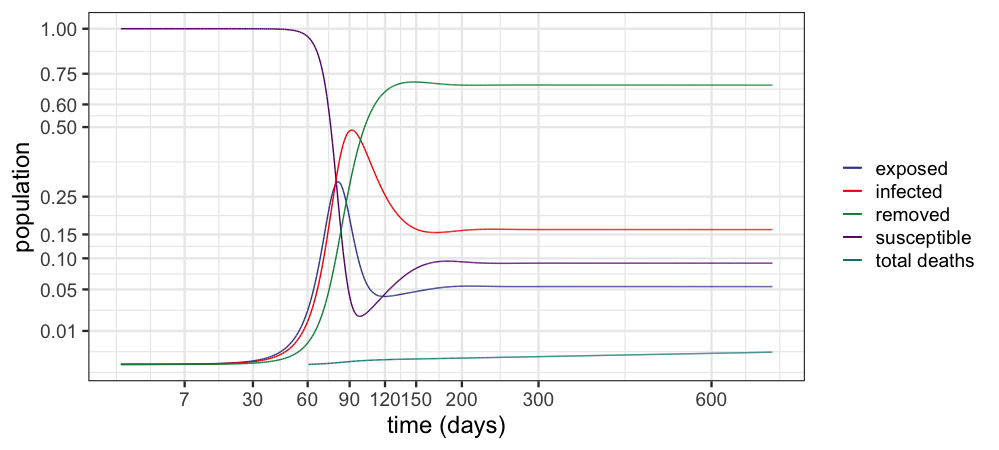}
	\caption{Dynamic behaviour of a single strain (\emph{Experiment 1}). \label{fig:single_strand} }
\end{figure}

Figure \ref{fig:single_strand} depicts the results for \emph{Experiment 1} and provides some insight into the long-term behaviour of the system with constant reinfection. For ease of interpretation, the population levels are shown as a proportion of the initial population $P(0)$ in this and all the remaining figures. Notice that the shares of susceptible, exposed, infected and removed (currently immune) individuals reach a sort of long-term equilibrium. For the specific parameters, the percentage of susceptible individuals stabilises just short of 10\%, whereas around 5\% of the individuals will be exposed - i.e. in the latency period - in the equilibrium. Furthermore, the level of infection in equilibrium is around 15\%, whereas around 60\% of the population will be intermittently immune to the virus in the long-term. Notice also that the number of deaths continues to increase over time.

We speculate that the level upon which the removed population stabilises in Figure \ref{fig:single_strand} provides some insight into the required levels of herd immunity for a given virus strain. In addition, the percentage of the population that is infected in the long-term, in the example around 15\% of the population at any given time, provides an insight into the burden of the epidemic on the health system. This level, coupled with the estimated number of cumulative deaths, may be used to inform healthcare policies in the long-term. Whereas the percentage of cumulative deaths is not to be viewed as an attempt to estimate such a level, given the uncertainty in the parameters and their variation around the world, it is provided here to offer some insights into the cumulative effect of the epidemic in the overall health of the population.

\begin{figure}[htbp!]
	\centering
	\includegraphics[width=\linewidth]{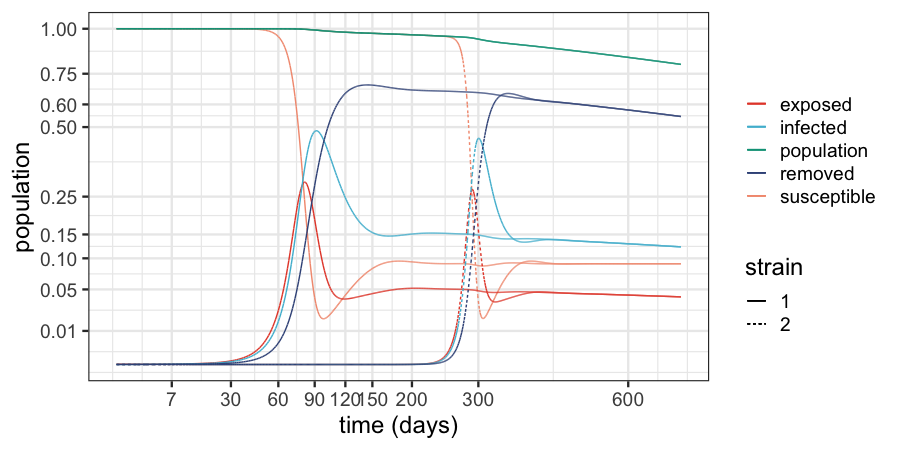}
	\caption{Dynamic behaviour of two strains (\emph{Experiment 2}). \label{fig:two_strains} }
\end{figure}

Figure \ref{fig:two_strains} illustrates the results for two viral strains (\emph{Experiment 2}), with the second appearing 180 days after the first. By second strain here we mean the first virus mutation that is sufficiently distinct as to not be affected by antibodies from previous strain. In this experiment, we assume that the parameters of the second strain are similar to those of the first.

Considering that the two strains are similar, the result in Figure  \ref{fig:two_strains} is quite intuitive. We observe that the second strain is simply a delayed version of the first outbreak, which makes sense given the similar parameters. The important feature here is that the second strain will add to the burden on the healthcare system, thereby increasing the levels of contamination and eventually doubling the burden. Notice, however, that at the peak of the second strain, most of the infections will be from this strain before the system eventually stabilises. Observe also the significant reduction of the population, which evinces a significant increase in deaths with respect to the single strain results. We argue that this should be considered to inform the decision makers. Indeed, strategies to prevent different strains from entering a given territory by enforcing testing upon arrival can be an important part of mitigation policies.

\begin{figure}[htbp!]
	\centering
	\includegraphics[width=\linewidth]{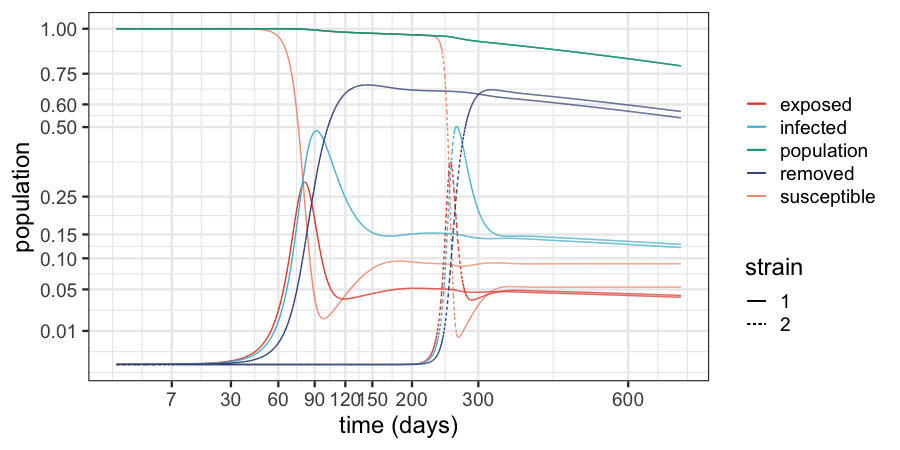}
	\caption{Dynamic behaviour of two strains, with second strain 70\% more transmissible (\emph{Experiment 3}). \label{fig:two_strains_70} }
\end{figure}

Inspired by the second strain reported in Britain, which is believe to be up to 70\% more transmissible than the first strain, the third experiment (\emph{Experiment 3}) replicates \emph{Experiment 2}, but considering a more transmissible second strain, with $\beta_2 = 1.7 \beta_1$. Figure \ref{fig:two_strains_70} depicts the results. One can notice that, as the second strain peaks, most of the new infections will be caused by this strain. As expected, one can also notice higher levels of mortality for the second strain, and a higher level of overall transmission of this strain as the system stabilises.

\subsection{Optimal mitigation strategies}

This section provides insights into the effect of the optimal control policy derived in Theorem \ref{thm:optimalcontrol} into the dynamics of the system over a two-year horizon. To simplify the results and facilitate the interpretation, we introduce a series of experiments with a single viral strain. This relies on the results of Section \ref{sec:stability} and Eq. \eqref{eq:R0}, which show that a control able to stabilise the most transmissible strain will also stabilise the remaining strains. Hence, an optimal policy derived for the single most transmissible strain can also be relied upon to stabilise the remaining strains.

We consider six sets of cost parameters $c_1$ and $c_2$ for the functional in Eq. \eqref{eq:functional} in the current experiment, which is labeled \emph{Experiment 4}. The parameters are described below:

\begin{center}
\begin{scriptsize}
\begin{tabular}{c|c|c|c|c|c|c}
\hline \hline
\textbf{Case} & A & B & C & D & E & F \\ \hline
\textbf{$c_1$} & 1 & 1 & 1 & 1 & 1 & 1 \\ \hline
\textbf{$c_2$} & $\text{ln}(\, P(0) \,) $ & $0.9 \,\text{ln}(\, P(0) \,)$ & $0.8 \, \text{ln}(\, P(0) \,)$ & $0.7 \, \text{ln}(\, P(0) \,)$ & $0.6 \, \text{ln}(\, P(0) \,)$ & $0.5 \, \text{ln}(\, P(0) \,) $ \\ \hline
\end{tabular}
\end{scriptsize}
\end{center}

\begin{figure}[htbp!]
	\centering
	\includegraphics[width=\linewidth]{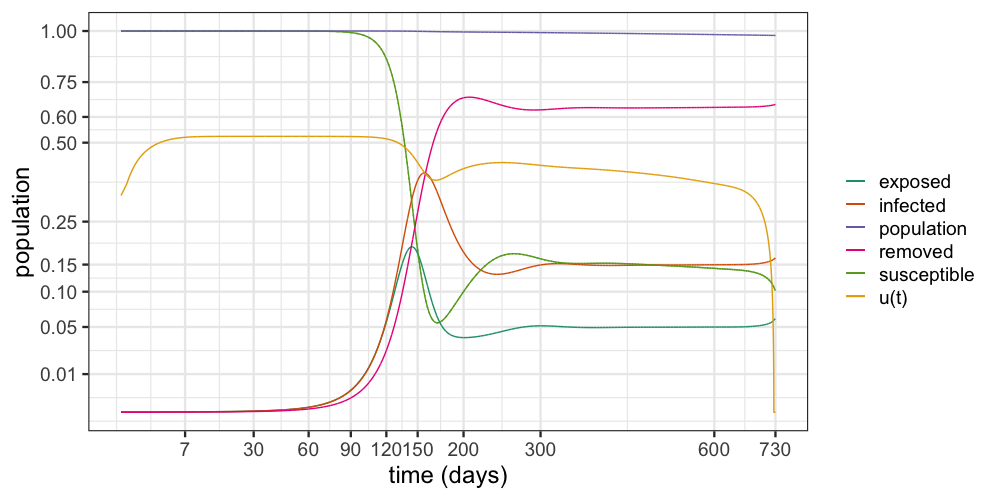}
	\caption{Optimal control policy for Case A \label{fig:caseA}}
\end{figure}

Cases A to F emulate a sequence of decreasing costs of mitigation measures, to provide an insight into the change in the optimal control as a function of such decrease. In Figure \ref{fig:caseA}, which depicts the results for Case A, one can notice that the optimal control stabilises around mitigating measures with a reduction of 50\% in transmission. After approximately four months the measures are slowly relaxed to yield a 37.5\% mitigation that produces a surge in infections. To respond to this surge, the mitigation is then restored to about 40\% and this level is decreased very slowly in the remaining horizon as the system stabilises.

\begin{figure}[htbp!]
	\centering
	\includegraphics[width=\linewidth]{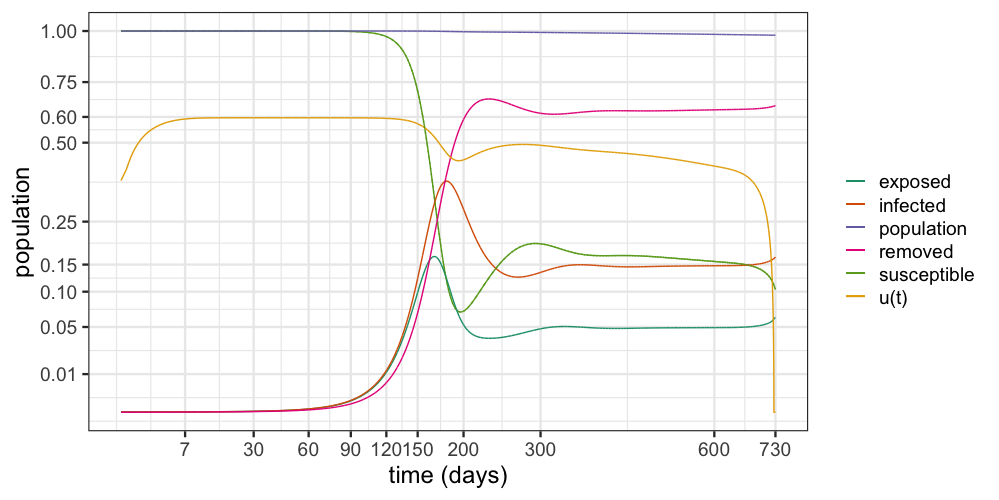}
	\caption{Optimal control policy for Case B \label{fig:caseB}}
\end{figure}

Case B introduces a small decrease in the cost of mitigation measures and one can see in Figure \ref{fig:caseB} that this results in an increase in the mitigation measures. It is also noteworthy that the overall behaviour of the mitigation measures $u(t)$ follows the same pattern as in the previous experiment.

\begin{figure}[htbp!]
	\centering
	\includegraphics[width=\linewidth]{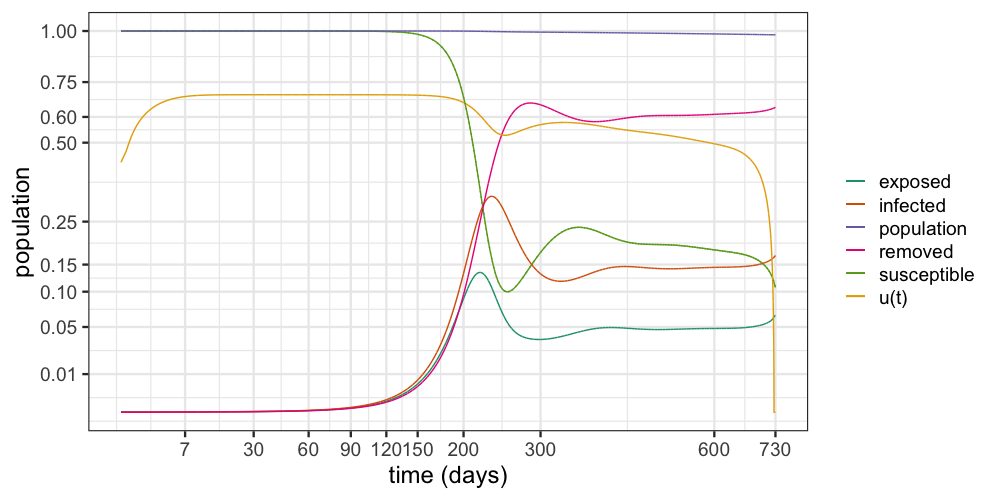}
	\caption{Optimal control policy for Case C \label{fig:caseC}}
\end{figure}

\begin{figure}[htbp!]
	\centering
	\includegraphics[width=\linewidth]{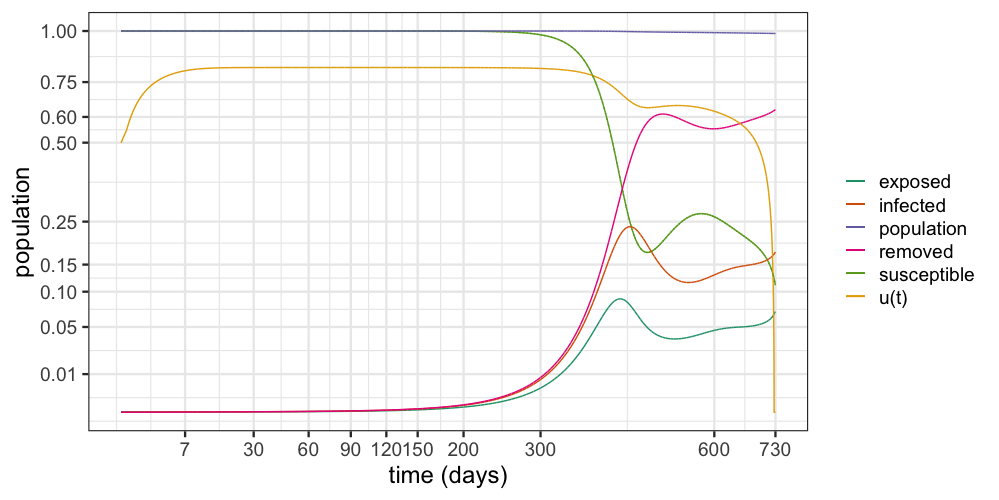}
	\caption{Optimal control policy for Case D\label{fig:caseD}}
\end{figure}

We also observe the same trends and overall behaviour of the mitigation measures in Cases C and D, see Figures \ref{fig:caseC}-\ref{fig:caseD}. However, as $c_2$ is decreased, we observe higher levels of mitigation measures over time. In addition, we can also observe that the relaxation of the mitigation measures is delayed with the increase of $c_2$.

When the cost of mitigation is sufficiently decreased, we observe that the mitigation measures tend to stabilise at a given level over the entire planning horizon. This happens in Cases E and F, depicted in Figures \ref{fig:caseE} and \ref{fig:caseF} below. The difference in these cases is that the mitigation levels stabilise around 80\% for Case E, while reaching around 88\% in Case F.

\begin{figure}[htbp!]
	\centering
	\includegraphics[width=\linewidth]{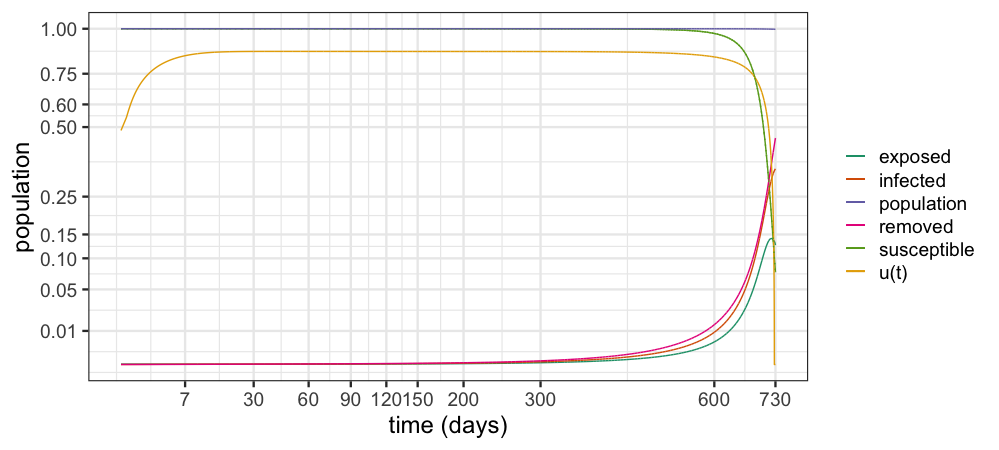}
	\caption{Optimal control policy for Case E \label{fig:caseE}}
\end{figure}

\begin{figure}[htbp!]
	\centering
	\includegraphics[width=\linewidth]{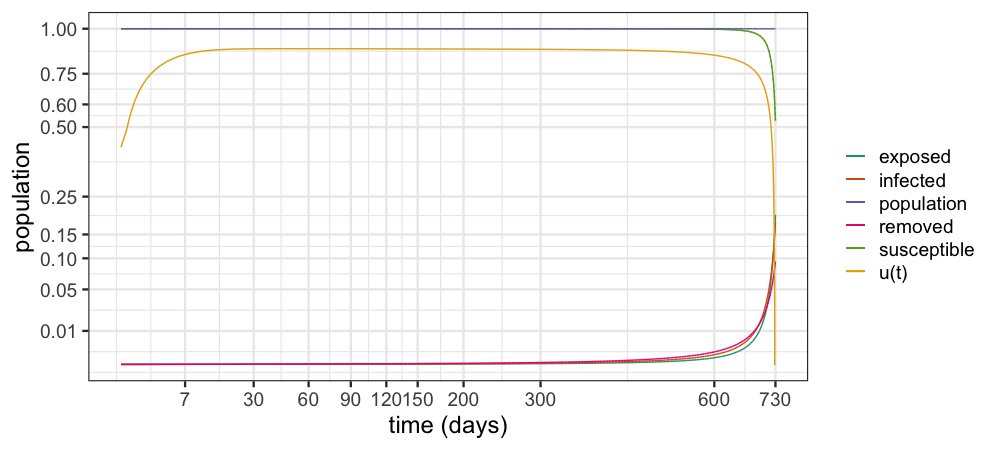}
	\caption{Optimal control policy for Case F \label{fig:caseF}}
\end{figure}

To sum up, the results illustrate the sensitivity of the model with respect to the perceived costs of mitigation measures. It is clear that different parameters may lead to highly distinct levels of infection and, consequently, overall deaths. Furthermore, in line with the results in \citep{Perkins2020}, we observe that large levels of control are needed from the outset to contain the infection levels and, therefore, preserve a wide range of feasible control measures as time elapses. On the other hand, if control measures are delayed, the burden on the healthcare system grows so rapidly that the decision makers may be left with no option but a full lock-down to curb the infection levels \citep{Tarrataca2021}.

\section{Concluding Remarks\label{sec:conclusion}}

This paper proposed a novel modelling framework based on the classical SEIR model that considers multiple viral strains, reinfections and optimal control. We  validated the framework in the light of the COVID-19 epidemic. The results are interpretable, robust and highlight a very intuitive result, namely that it is possible to contain the epidemic by focusing on the most transmissible strain.

The results show that, in the absence of mitigating measures, an epidemic with a single viral strain and reinfection will reach an equilibrium after the peak of infections. Whilst real-world data suggest that the peak is not manageable by any healthcare system in the world, it is evident that even the equilibrium may imply levels of infection that will challenge healthcare resources in many regions of the world.

The results also suggest that, with insufficient mitigation measures, an epidemic with a second wave includes a second peak of infections that is higher than the first if the second strain is as transmissible as the first; this is because the second peak combines infections from both strains. Moreover, the number of deaths increases considerably in the presence of the second strain. In view of the second COVID-19 strain in the UK, we also simulated a scenario where the second strain is 70\% more transmissible than the first. This causes a larger and sharper peak in the second wave, in which the majority of cases will belong to the second strain not just because it is more transmissible, but also because it peaks when the first strain had already stabilised. The results are consistent with the reports of the second strain in the UK and in the state of Amazonas, Brazil, where the second wave caused an explosive increase in new cases, as well as a complete depletion of healthcare resources.

Finally, we propose and solve an optimal control problem to derive optimal mitigation measures whilst considering that the cost of mitigation measures grows exponentially as a function of the mitigation effort. Because controlling the single most transmissible strain suffices, we derived optimal control strategies for a single strain to facilitate the interpretation of the results. Consistently with previous literature, our findings suggest that an early relaxation of lockdown measures causes a sharp increase in the number of new cases which, in turn, will lead to the need for more restrictive measures. As expected, the optimal levels of isolation are very sensitive to the mitigation costs, with lower costs resulting in higher levels of restrictive measures. Furthermore, only persistently high levels of mitigation measures are capable of containing the epidemic in the long-term, which suggests that an efficient deployment of mass vaccination is the single most effective way out of the current epidemic.

\section*{Acknowledgements}

This study was partly supported by the Brazilian Research Council—CNPq, under grant \#311075/2018-5 and by Coordenação de Aperfeiçoamento de Pessoal de Nível Superior—Brasil (CAPES) [Finance Code 001]


\begin{thebibliography}{39}
\expandafter\ifx\csname natexlab\endcsname\relax\def\natexlab#1{#1}\fi
\providecommand{\url}[1]{\texttt{#1}}
\providecommand{\href}[2]{#2}
\providecommand{\path}[1]{#1}
\providecommand{\DOIprefix}{doi:}
\providecommand{\ArXivprefix}{arXiv:}
\providecommand{\URLprefix}{URL: }
\providecommand{\Pubmedprefix}{pmid:}
\providecommand{\doi}[1]{\href{http://dx.doi.org/#1}{\path{#1}}}
\providecommand{\Pubmed}[1]{\href{pmid:#1}{\path{#1}}}
\providecommand{\bibinfo}[2]{#2}
\ifx\xfnm\relax \def\xfnm[#1]{\unskip,\space#1}\fi
\bibitem[{Bacaer(2011)}]{Kermack}
\bibinfo{author}{Bacaer, N.}, \bibinfo{year}{2011}.
\newblock \bibinfo{title}{{McKendrick and Kermack} on epidemic modelling
  (1926–1927)}.
\newblock \bibinfo{edition}{1st} ed., \bibinfo{publisher}{Springer},
  \bibinfo{address}{Lodon}.
\bibitem[{Bentaleb et~al.(2020)Bentaleb, Harroudi, Amine and
  Allali}]{Bentaleb2020}
\bibinfo{author}{Bentaleb, D.}, \bibinfo{author}{Harroudi, S.},
  \bibinfo{author}{Amine, S.}, \bibinfo{author}{Allali, K.},
  \bibinfo{year}{2020}.
\newblock \bibinfo{title}{Analysis and optimal control of a multistrain {SEIR}
  epidemic model with saturated incidence rate and treatment}.
\newblock \bibinfo{journal}{Differential Equations and Dynamical Systems}
  \DOIprefix\doi{10.1007/s12591-020-00544-6}.
\bibitem[{Bonifacio et~al.(2020)Bonifacio, Pereira, Araujo, Balbao, Fonseca,
  Passos and Bellissimo-Rodrigues}]{BONIFACIO2020}
\bibinfo{author}{Bonifacio, L.P.}, \bibinfo{author}{Pereira, A.P.S.},
  \bibinfo{author}{Araujo, D.C.A.}, \bibinfo{author}{Balbao, V.M.P.},
  \bibinfo{author}{Fonseca, B.A.L.}, \bibinfo{author}{Passos, A.D.C.},
  \bibinfo{author}{Bellissimo-Rodrigues, F.}, \bibinfo{year}{2020}.
\newblock \bibinfo{title}{{Are {SARS-CoV-2} reinfection and Covid-19 recurrence
  possible? a case report from Brazil}}.
\newblock \bibinfo{journal}{Revista da Sociedade Brasileira de Medicina
  Tropical} \bibinfo{volume}{53}.
\newblock \DOIprefix\doi{10.1590/0037-8682-0619-2020}.
\bibitem[{Bryson and Ho(1970)}]{Bryson1975}
\bibinfo{author}{Bryson, A.E.}, \bibinfo{author}{Ho, Y.C.},
  \bibinfo{year}{1970}.
\newblock \bibinfo{title}{Applied Optimal Control: {Optimization}, Estimation,
  and Control}.
\newblock \bibinfo{publisher}{CRC Press}.
\newblock \DOIprefix\doi{10.1201/9781315137667}.
\bibitem[{Bursac et~al.(2020)Bursac, Madubueze, Dachollom and
  Onwubuya}]{Bursac2020}
\bibinfo{author}{Bursac, Z.}, \bibinfo{author}{Madubueze, C.E.},
  \bibinfo{author}{Dachollom, S.}, \bibinfo{author}{Onwubuya, I.O.},
  \bibinfo{year}{2020}.
\newblock \bibinfo{title}{Controlling the spread of {COVID-19: Optimal} control
  analysis}.
\newblock \bibinfo{journal}{Computational and Mathematical Methods in Medicine}
  \bibinfo{volume}{2020}, \bibinfo{pages}{6862516}.
\newblock \DOIprefix\doi{10.1155/2020/6862516}.
\bibitem[{Callaway(2020)}]{Callaway2020}
\bibinfo{author}{Callaway, E.}, \bibinfo{year}{2020}.
\newblock \bibinfo{title}{The coronavirus is mutating — does it matter?}
\newblock \bibinfo{journal}{Nature} \bibinfo{volume}{585},
  \bibinfo{pages}{174--177}.
\newblock \DOIprefix\doi{10.1038/d41586-020-02544-6}.
\bibitem[{Candido et~al.(2020)Candido, Claro, Jesus, Souza, Moreira,
  Simon~Dellicour, Mellan, du~Plessis, Pereira, Sales, Erika R.~Manuli,
  Thézé, Almeida, Menezes, Voloch, Fumagalli, Coletti, da~Silva, Ramundo,
  Amorim, Hoeltgebaum, Mishra, Gill, Carvalho, Buss, Prete~Jr., Ashworth,
  Nakaya, Peixoto, Brady, Nicholls, Tanuri, Rossi, Braga, Gerber, Guimarães,
  Gaburo~Jr., Alencar, Ferreira, Lima, Levi, Granato, Ferreira, Francisco~Jr.,
  Granja, Garcia, Moretti, Perroud~Jr., Castiñeiras, Lazari, Hill, Santos,
  Simeoni, Forato, Sposito, Schreiber, Santos, de~Sá, Souza, Resende-Moreira,
  Teixeira, Hubner, Leme, Moreira, Nogueira, {Brazil-UK Centre for Arbovirus
  Discovery, Diagnosis, Genomics and Epidemiology (CADDE) Genomic Network},
  Ferguson, Costa, Proenca-Modena, Vasconcelos, Bhatt, Lemey, Wu, Rambaut,
  Loman, Aguiar, Pybus, Sabino and Faria}]{anatereza}
\bibinfo{author}{Candido, D.S.}, \bibinfo{author}{Claro, I.M.},
  \bibinfo{author}{Jesus, J.G.}, \bibinfo{author}{Souza, W.M.},
  \bibinfo{author}{Moreira, F.R.R.}, \bibinfo{author}{Simon~Dellicour, S.},
  \bibinfo{author}{Mellan, T.A.}, \bibinfo{author}{du~Plessis, L.},
  \bibinfo{author}{Pereira, R.H.M.}, \bibinfo{author}{Sales, F.C.S.},
  \bibinfo{author}{Erika R.~Manuli, E.R.}, \bibinfo{author}{Thézé, J.},
  \bibinfo{author}{Almeida, L.}, \bibinfo{author}{Menezes, M.T.},
  \bibinfo{author}{Voloch, C.M.}, \bibinfo{author}{Fumagalli, M.J.},
  \bibinfo{author}{Coletti, T.M.}, \bibinfo{author}{da~Silva, C.A.M.},
  \bibinfo{author}{Ramundo, M.S.}, \bibinfo{author}{Amorim, M.R.},
  \bibinfo{author}{Hoeltgebaum, H.H.}, \bibinfo{author}{Mishra, S.},
  \bibinfo{author}{Gill, M.S.}, \bibinfo{author}{Carvalho, L.M.},
  \bibinfo{author}{Buss, L.F.}, \bibinfo{author}{Prete~Jr., C.A.},
  \bibinfo{author}{Ashworth, J.}, \bibinfo{author}{Nakaya, H.I.},
  \bibinfo{author}{Peixoto, P.S.}, \bibinfo{author}{Brady, O.J.},
  \bibinfo{author}{Nicholls, S.M.}, \bibinfo{author}{Tanuri, A.},
  \bibinfo{author}{Rossi, A.D.}, \bibinfo{author}{Braga, C.K.V.},
  \bibinfo{author}{Gerber, A.L.}, \bibinfo{author}{Guimarães, A.P.C.},
  \bibinfo{author}{Gaburo~Jr., N.}, \bibinfo{author}{Alencar, S.},
  \bibinfo{author}{Ferreira, A.C.S.}, \bibinfo{author}{Lima, C.X.},
  \bibinfo{author}{Levi, J.E.}, \bibinfo{author}{Granato, C.},
  \bibinfo{author}{Ferreira, G.M.}, \bibinfo{author}{Francisco~Jr., R.S.},
  \bibinfo{author}{Granja, F.}, \bibinfo{author}{Garcia, M.T.},
  \bibinfo{author}{Moretti, M.L.}, \bibinfo{author}{Perroud~Jr., M.W.},
  \bibinfo{author}{Castiñeiras, T.M.P.P.}, \bibinfo{author}{Lazari, C.S.},
  \bibinfo{author}{Hill, S.C.}, \bibinfo{author}{Santos, A.A.S.},
  \bibinfo{author}{Simeoni, C.L.}, \bibinfo{author}{Forato, J.},
  \bibinfo{author}{Sposito, A.C.}, \bibinfo{author}{Schreiber, A.Z.},
  \bibinfo{author}{Santos, M.N.N.}, \bibinfo{author}{de~Sá, C.Z.},
  \bibinfo{author}{Souza, R.P.}, \bibinfo{author}{Resende-Moreira, L.C.},
  \bibinfo{author}{Teixeira, M.M.}, \bibinfo{author}{Hubner, J.},
  \bibinfo{author}{Leme, P.A.F.}, \bibinfo{author}{Moreira, R.G.},
  \bibinfo{author}{Nogueira, M.L.}, \bibinfo{author}{{Brazil-UK Centre for
  Arbovirus Discovery, Diagnosis, Genomics and Epidemiology (CADDE) Genomic
  Network}}, \bibinfo{author}{Ferguson, N.M.}, \bibinfo{author}{Costa, S.F.},
  \bibinfo{author}{Proenca-Modena, J.L.}, \bibinfo{author}{Vasconcelos,
  A.T.R.}, \bibinfo{author}{Bhatt, S.}, \bibinfo{author}{Lemey, P.},
  \bibinfo{author}{Wu, C.H.}, \bibinfo{author}{Rambaut, A.},
  \bibinfo{author}{Loman, N.J.}, \bibinfo{author}{Aguiar, R.S.},
  \bibinfo{author}{Pybus, O.G.}, \bibinfo{author}{Sabino, E.C.},
  \bibinfo{author}{Faria, N.R.}, \bibinfo{year}{2020}.
\newblock \bibinfo{title}{Evolution and epidemic spread of {SARS-CoV-2} in
  {Brazil}}.
\newblock \bibinfo{journal}{Science} \bibinfo{volume}{369},
  \bibinfo{pages}{1255--1260}.
\newblock \DOIprefix\doi{10.1126/science.abd2161}.
\bibitem[{Dan et~al.(2021)Dan, Mateus, Kato, Hastie, Yu, Faliti, Grifoni,
  Ramirez, Haupt, Frazier, Nakao, Rayaprolu, Rawlings, Peters, Krammer, Simon,
  Saphire, Smith, Weiskopf, Sette and Crotty}]{Daneabf4063}
\bibinfo{author}{Dan, J.M.}, \bibinfo{author}{Mateus, J.},
  \bibinfo{author}{Kato, Y.}, \bibinfo{author}{Hastie, K.M.},
  \bibinfo{author}{Yu, E.D.}, \bibinfo{author}{Faliti, C.E.},
  \bibinfo{author}{Grifoni, A.}, \bibinfo{author}{Ramirez, S.I.},
  \bibinfo{author}{Haupt, S.}, \bibinfo{author}{Frazier, A.},
  \bibinfo{author}{Nakao, C.}, \bibinfo{author}{Rayaprolu, V.},
  \bibinfo{author}{Rawlings, S.A.}, \bibinfo{author}{Peters, B.},
  \bibinfo{author}{Krammer, F.}, \bibinfo{author}{Simon, V.},
  \bibinfo{author}{Saphire, E.O.}, \bibinfo{author}{Smith, D.M.},
  \bibinfo{author}{Weiskopf, D.}, \bibinfo{author}{Sette, A.},
  \bibinfo{author}{Crotty, S.}, \bibinfo{year}{2021}.
\newblock \bibinfo{title}{Immunological memory to {SARS-CoV-2} assessed for up
  to 8 months after infection}.
\newblock \bibinfo{journal}{Science} \DOIprefix\doi{10.1126/science.abf4063}.
\bibitem[{Dawood(2020)}]{DAWOOD2020100673}
\bibinfo{author}{Dawood, A.}, \bibinfo{year}{2020}.
\newblock \bibinfo{title}{Mutated {COVID-19} may foretell a great risk for
  mankind in the future}.
\newblock \bibinfo{journal}{New Microbes and New Infections}
  \bibinfo{volume}{35}, \bibinfo{pages}{100673}.
\newblock \DOIprefix\doi{10.1016/j.nmni.2020.100673}.
\bibitem[{Edridge et~al.(2020)Edridge, Kaczorowska, Hoste, Bakker, Klein,
  Loens, Jebbink, Matser, Kinsella, Rueda, Ieven, Goossens, Prins, Sastre,
  Deijs and van~der Hoek}]{Edridge_2020}
\bibinfo{author}{Edridge, A.W.D.}, \bibinfo{author}{Kaczorowska, J.},
  \bibinfo{author}{Hoste, A.C.R.}, \bibinfo{author}{Bakker, M.},
  \bibinfo{author}{Klein, M.}, \bibinfo{author}{Loens, K.},
  \bibinfo{author}{Jebbink, M.F.}, \bibinfo{author}{Matser, A.},
  \bibinfo{author}{Kinsella, C.M.}, \bibinfo{author}{Rueda, P.},
  \bibinfo{author}{Ieven, M.}, \bibinfo{author}{Goossens, H.},
  \bibinfo{author}{Prins, M.}, \bibinfo{author}{Sastre, P.},
  \bibinfo{author}{Deijs, M.}, \bibinfo{author}{van~der Hoek, L.},
  \bibinfo{year}{2020}.
\newblock \bibinfo{title}{Seasonal coronavirus protective immunity is
  short-lasting}.
\newblock \bibinfo{journal}{Nature Medicine} \bibinfo{volume}{26},
  \bibinfo{pages}{1691--1693}.
\newblock \DOIprefix\doi{10.1038/s41591-020-1083-1}.
\bibitem[{Etbaigha et~al.(2018)Etbaigha, Willms and Poljak}]{Etbaigha2018}
\bibinfo{author}{Etbaigha, F.}, \bibinfo{author}{Willms, A.R.},
  \bibinfo{author}{Poljak, Z.}, \bibinfo{year}{2018}.
\newblock \bibinfo{title}{An {SEIR} model of influenza {A} virus infection and
  reinfection within a farrow-to-finish swine farm}.
\newblock \bibinfo{journal}{PLOS ONE} \bibinfo{volume}{13},
  \bibinfo{pages}{1--19}.
\newblock \DOIprefix\doi{10.1371/journal.pone.0202493}.
\bibitem[{Ferguson et~al.(2020)Ferguson, Laydon, Nedjati~Gilani, Imai, Ainslie,
  Baguelin, Bhatia, Boonyasiri, Cucunuba~Perez, Cuomo-Dannenburg, Dighe,
  Dorigatti, Fu, Gaythorpe, Green, Hamlet, Hinsley, Okell, Van~Elsland,
  Thompson, Verity, Volz, Wang, Wang, Walker, Winskill, Whittaker, Donnelly,
  Riley and Ghani}]{ferguson2020}
\bibinfo{author}{Ferguson, N.}, \bibinfo{author}{Laydon, D.},
  \bibinfo{author}{Nedjati~Gilani, G.}, \bibinfo{author}{Imai, N.},
  \bibinfo{author}{Ainslie, K.}, \bibinfo{author}{Baguelin, M.},
  \bibinfo{author}{Bhatia, S.}, \bibinfo{author}{Boonyasiri, A.},
  \bibinfo{author}{Cucunuba~Perez, Z.}, \bibinfo{author}{Cuomo-Dannenburg, G.},
  \bibinfo{author}{Dighe, A.}, \bibinfo{author}{Dorigatti, I.},
  \bibinfo{author}{Fu, H.}, \bibinfo{author}{Gaythorpe, K.},
  \bibinfo{author}{Green, W.}, \bibinfo{author}{Hamlet, A.},
  \bibinfo{author}{Hinsley, W.}, \bibinfo{author}{Okell, L.},
  \bibinfo{author}{Van~Elsland, S.}, \bibinfo{author}{Thompson, H.},
  \bibinfo{author}{Verity, R.}, \bibinfo{author}{Volz, E.},
  \bibinfo{author}{Wang, H.}, \bibinfo{author}{Wang, Y.},
  \bibinfo{author}{Walker, P.}, \bibinfo{author}{Winskill, P.},
  \bibinfo{author}{Whittaker, C.}, \bibinfo{author}{Donnelly, C.},
  \bibinfo{author}{Riley, S.}, \bibinfo{author}{Ghani, A.},
  \bibinfo{year}{2020}.
\newblock \bibinfo{title}{Report 9: Impact of non-pharmaceutical interventions
  {(NPIs)} to reduce {COVID-19} mortality and healthcare demand}.
\newblock \bibinfo{type}{Technical Report}. Imperial College London.
\newblock \DOIprefix\doi{10.25561/77482}.
\bibitem[{Flaxman et~al.(2020)Flaxman, Mishra, Gandy, Unwin, Coupland, Mellan,
  Zhu, Berah, Eaton, Guzman, Schmit, Cilloni, Ainslie, Baguelin, Blake,
  Boonyasiri, Boyd, Cattarino, Ciavarella, Cooper, Perez, Cuomo-Dannenburg,
  Dighe, Djaafara, Dorigatti, {Van Elsland}, Fitzjohn, Fu, Gaythorpe,
  Geidelberg, Grassly, Green, Hallett, Hamlet, Hinsley, Jeffrey, Jorgensen,
  Knock, Laydon, Gilani, Nouvellet, Parag, Siveroni, Thompson, Verity, Volz,
  Walters, Wang, Wang, Watson, Winskill, Xi, Whittaker, Walker, Ghani,
  Donnelly, Riley, Okell, Vollmer, Ferguson and Bhatt}]{flaxman2020}
\bibinfo{author}{Flaxman, S.}, \bibinfo{author}{Mishra, S.},
  \bibinfo{author}{Gandy, A.}, \bibinfo{author}{Unwin, H.},
  \bibinfo{author}{Coupland, H.}, \bibinfo{author}{Mellan, T.},
  \bibinfo{author}{Zhu, H.}, \bibinfo{author}{Berah, T.},
  \bibinfo{author}{Eaton, J.}, \bibinfo{author}{Guzman, P.P.},
  \bibinfo{author}{Schmit, N.}, \bibinfo{author}{Cilloni, L.},
  \bibinfo{author}{Ainslie, K.}, \bibinfo{author}{Baguelin, M.},
  \bibinfo{author}{Blake, I.}, \bibinfo{author}{Boonyasiri, A.},
  \bibinfo{author}{Boyd, O.}, \bibinfo{author}{Cattarino, L.},
  \bibinfo{author}{Ciavarella, C.}, \bibinfo{author}{Cooper, L.},
  \bibinfo{author}{Perez, Z.C.}, \bibinfo{author}{Cuomo-Dannenburg, G.},
  \bibinfo{author}{Dighe, A.}, \bibinfo{author}{Djaafara, A.},
  \bibinfo{author}{Dorigatti, I.}, \bibinfo{author}{{Van Elsland}, S.},
  \bibinfo{author}{Fitzjohn, R.}, \bibinfo{author}{Fu, H.},
  \bibinfo{author}{Gaythorpe, K.}, \bibinfo{author}{Geidelberg, L.},
  \bibinfo{author}{Grassly, N.}, \bibinfo{author}{Green, W.},
  \bibinfo{author}{Hallett, T.}, \bibinfo{author}{Hamlet, A.},
  \bibinfo{author}{Hinsley, W.}, \bibinfo{author}{Jeffrey, B.},
  \bibinfo{author}{Jorgensen, D.}, \bibinfo{author}{Knock, E.},
  \bibinfo{author}{Laydon, D.}, \bibinfo{author}{Gilani, G.N.},
  \bibinfo{author}{Nouvellet, P.}, \bibinfo{author}{Parag, K.},
  \bibinfo{author}{Siveroni, I.}, \bibinfo{author}{Thompson, H.},
  \bibinfo{author}{Verity, R.}, \bibinfo{author}{Volz, E.},
  \bibinfo{author}{Walters, C.}, \bibinfo{author}{Wang, H.},
  \bibinfo{author}{Wang, Y.}, \bibinfo{author}{Watson, O.},
  \bibinfo{author}{Winskill, P.}, \bibinfo{author}{Xi, X.},
  \bibinfo{author}{Whittaker, C.}, \bibinfo{author}{Walker, P.},
  \bibinfo{author}{Ghani, A.}, \bibinfo{author}{Donnelly, C.},
  \bibinfo{author}{Riley, S.}, \bibinfo{author}{Okell, L.},
  \bibinfo{author}{Vollmer, M.}, \bibinfo{author}{Ferguson, N.},
  \bibinfo{author}{Bhatt, S.}, \bibinfo{year}{2020}.
\newblock \bibinfo{title}{Report 13: Estimating the number of infections and
  the impact of non-pharmaceutical interventions on {COVID-19} in 11 {European}
  countries}.
\newblock \bibinfo{type}{Technical Report}. Imperial College London.
\newblock \DOIprefix\doi{10.25561/77731}.
\bibitem[{Frid et~al.(2003)Frid, Jabin and Perthame}]{Perthame2003}
\bibinfo{author}{Frid, H.}, \bibinfo{author}{Jabin, P.E.},
  \bibinfo{author}{Perthame, B.}, \bibinfo{year}{2003}.
\newblock \bibinfo{title}{Global stability of steady solutions for a model in
  virus dynamics}.
\newblock \bibinfo{journal}{ESAIM: Mathematical Modelling and Numerical
  Analysis} \bibinfo{volume}{37}, \bibinfo{pages}{709--723}.
\newblock \DOIprefix\doi{10.1051/m2an:2003045}.
\bibitem[{Fudolig and Howard(2020)}]{Fudolig2020}
\bibinfo{author}{Fudolig, M.}, \bibinfo{author}{Howard, R.},
  \bibinfo{year}{2020}.
\newblock \bibinfo{title}{The local stability of a modified multi-strain {SIR}
  model for emerging viral strains}.
\newblock \bibinfo{journal}{PLoS ONE} \bibinfo{volume}{12},
  \bibinfo{pages}{e0243408}.
\newblock \DOIprefix\doi{10.1371/journal.pone.0243408}.
\bibitem[{Gubar et~al.(2017)Gubar, Zhu and Taynitskiy}]{Gubar2017}
\bibinfo{author}{Gubar, E.}, \bibinfo{author}{Zhu, Q.},
  \bibinfo{author}{Taynitskiy, V.}, \bibinfo{year}{2017}.
\newblock \bibinfo{title}{Optimal control of multi-strain epidemic processes in
  complex networks}, in: \bibinfo{editor}{Duan, L.}, \bibinfo{editor}{Sanjab,
  A.}, \bibinfo{editor}{Li, H.}, \bibinfo{editor}{Chen, X.},
  \bibinfo{editor}{Materassi, D.}, \bibinfo{editor}{Elazouzi, R.} (Eds.),
  \bibinfo{booktitle}{Game Theory for Networks}, \bibinfo{publisher}{Springer
  International Publishing}, \bibinfo{address}{Cham}. pp.
  \bibinfo{pages}{108--117}.
\bibitem[{Kantner and Koprucki(2020)}]{Kantner2020}
\bibinfo{author}{Kantner, M.}, \bibinfo{author}{Koprucki, T.},
  \bibinfo{year}{2020}.
\newblock \bibinfo{title}{Beyond just ``flattening the curve'': Optimal control
  of epidemics with purely non-pharmaceutical interventions}.
\newblock \bibinfo{journal}{Journal of Mathematics in Industry}
  \bibinfo{volume}{10}, \bibinfo{pages}{23}.
\newblock \DOIprefix\doi{10.1186/s13362-020-00091-3}.
\bibitem[{Kermack et~al.(1927)Kermack, McKndrick and Walker}]{Kermack1927}
\bibinfo{author}{Kermack, W.O.}, \bibinfo{author}{McKndrick, A.G.},
  \bibinfo{author}{Walker, G.T.}, \bibinfo{year}{1927}.
\newblock \bibinfo{title}{A contribution to the mathematical theory of
  epidemics}.
\newblock \bibinfo{journal}{Proceedings of the Royal Society of London. Series
  A, Containing Papers of a Mathematical and Physical Character}
  \bibinfo{volume}{115}, \bibinfo{pages}{700--721}.
\newblock \DOIprefix\doi{10.1098/rspa.1927.0118}.
\bibitem[{Khyar and Allali(2020)}]{Khyar2020}
\bibinfo{author}{Khyar, O.}, \bibinfo{author}{Allali, K.},
  \bibinfo{year}{2020}.
\newblock \bibinfo{title}{Global dynamics of a multi-strain {SEIR} epidemic
  model with general incidence rates: application to {COVID-19} pandemic}.
\newblock \bibinfo{journal}{Nonlinear Dyn} \bibinfo{volume}{102},
  \bibinfo{pages}{489--509}.
\newblock \DOIprefix\doi{10.1007/s11071-020-05929-4}.
\bibitem[{Kirby(2021)}]{KIRBY2021}
\bibinfo{author}{Kirby, T.}, \bibinfo{year}{2021}.
\newblock \bibinfo{title}{New variant of sars-cov-2 in uk causes surge of
  covid-19}.
\newblock \bibinfo{journal}{The Lancet Respiratory Medicine}
  \DOIprefix\doi{10.1016/S2213-2600(21)00005-9}.
\bibitem[{Kirk(1970)}]{Kirk1970}
\bibinfo{author}{Kirk, D.}, \bibinfo{year}{1970}.
\newblock \bibinfo{title}{Optimal Control Theory: {An} Introduction}.
\newblock Networks Series, \bibinfo{publisher}{Prentice-Hall}.
\bibitem[{Korber et~al.(2020)Korber, Fischer, Gnanakaran, Yoon, Theiler,
  Abfalterer, Hengartner, Giorgi, Bhattacharya, Foley, Hastie, Parker,
  Partridge, Evans, Freeman, {de Silva}, Angyal, Brown, Carrilero, Green,
  Groves, Johnson, Keeley, Lindsey, Parsons, Raza, Rowland-Jones, Smith,
  Tucker, Wang, Wyles, McDanal, Perez, Tang, Moon-Walker, Whelan, LaBranche,
  Saphire and Montefiori}]{KORBER2020812}
\bibinfo{author}{Korber, B.}, \bibinfo{author}{Fischer, W.M.},
  \bibinfo{author}{Gnanakaran, S.}, \bibinfo{author}{Yoon, H.},
  \bibinfo{author}{Theiler, J.}, \bibinfo{author}{Abfalterer, W.},
  \bibinfo{author}{Hengartner, N.}, \bibinfo{author}{Giorgi, E.E.},
  \bibinfo{author}{Bhattacharya, T.}, \bibinfo{author}{Foley, B.},
  \bibinfo{author}{Hastie, K.M.}, \bibinfo{author}{Parker, M.D.},
  \bibinfo{author}{Partridge, D.G.}, \bibinfo{author}{Evans, C.M.},
  \bibinfo{author}{Freeman, T.M.}, \bibinfo{author}{{de Silva}, T.I.},
  \bibinfo{author}{Angyal, A.}, \bibinfo{author}{Brown, R.L.},
  \bibinfo{author}{Carrilero, L.}, \bibinfo{author}{Green, L.R.},
  \bibinfo{author}{Groves, D.C.}, \bibinfo{author}{Johnson, K.J.},
  \bibinfo{author}{Keeley, A.J.}, \bibinfo{author}{Lindsey, B.B.},
  \bibinfo{author}{Parsons, P.J.}, \bibinfo{author}{Raza, M.},
  \bibinfo{author}{Rowland-Jones, S.}, \bibinfo{author}{Smith, N.},
  \bibinfo{author}{Tucker, R.M.}, \bibinfo{author}{Wang, D.},
  \bibinfo{author}{Wyles, M.D.}, \bibinfo{author}{McDanal, C.},
  \bibinfo{author}{Perez, L.G.}, \bibinfo{author}{Tang, H.},
  \bibinfo{author}{Moon-Walker, A.}, \bibinfo{author}{Whelan, S.P.},
  \bibinfo{author}{LaBranche, C.C.}, \bibinfo{author}{Saphire, E.O.},
  \bibinfo{author}{Montefiori, D.C.}, \bibinfo{year}{2020}.
\newblock \bibinfo{title}{Tracking changes in {SARS-CoV-2} spike: {Evidence
  that D614G} increases infectivity of the {COVID-19} virus}.
\newblock \bibinfo{journal}{Cell} \bibinfo{volume}{182}, \bibinfo{pages}{812 --
  827.e19}.
\newblock \DOIprefix\doi{10.1016/j.cell.2020.06.043}.
\bibitem[{Long et~al.(2020)Long, Tang, Shi, Li, Deng, Yuan, Hu, Xu, Zhang, Lv,
  Su, Zhang, Gong, Wu, Liu, Li, Qiu, Chen and Huang}]{long_2020}
\bibinfo{author}{Long, Q.X.}, \bibinfo{author}{Tang, X.J.},
  \bibinfo{author}{Shi, Q.L.}, \bibinfo{author}{Li, Q.}, \bibinfo{author}{Deng,
  H.J.}, \bibinfo{author}{Yuan, J.}, \bibinfo{author}{Hu, J.L.},
  \bibinfo{author}{Xu, W.}, \bibinfo{author}{Zhang, Y.}, \bibinfo{author}{Lv,
  F.J.}, \bibinfo{author}{Su, K.}, \bibinfo{author}{Zhang, F.},
  \bibinfo{author}{Gong, J.}, \bibinfo{author}{Wu, B.}, \bibinfo{author}{Liu,
  X.M.}, \bibinfo{author}{Li, J.J.}, \bibinfo{author}{Qiu, J.F.},
  \bibinfo{author}{Chen, J.}, \bibinfo{author}{Huang, A.L.},
  \bibinfo{year}{2020}.
\newblock \bibinfo{title}{Clinical and immunological assessment of asymptomatic
  sars-cov-2 infections}.
\newblock \bibinfo{journal}{Nature Medicine} \bibinfo{volume}{26},
  \bibinfo{pages}{1200--1204}.
\newblock \DOIprefix\doi{10.1038/s41591-020-0965-6}.
\bibitem[{McMahon and Robb(2020)}]{McMahon2020}
\bibinfo{author}{McMahon, A.}, \bibinfo{author}{Robb, N.C.},
  \bibinfo{year}{2020}.
\newblock \bibinfo{title}{Reinfection with {SARS-CoV-2}: {Discrete SIR}
  (susceptible, infected, recovered) modeling using empirical infection data}.
\newblock \bibinfo{journal}{JMIR public health and surveillance}
  \bibinfo{volume}{6}, \bibinfo{pages}{e21168--e21168}.
\newblock \DOIprefix\doi{10.2196/21168}.
\bibitem[{Nonaka et~al.(2021)Nonaka, Franco, Graf, A.~V. A.~Mendes, Giovanetti
  and Souza}]{Nonaka2021}
\bibinfo{author}{Nonaka, C.K.V.}, \bibinfo{author}{Franco, M.M.},
  \bibinfo{author}{Graf, T.}, \bibinfo{author}{A.~V. A.~Mendes, A.R.S.A.},
  \bibinfo{author}{Giovanetti, M.}, \bibinfo{author}{Souza, B.S.F.},
  \bibinfo{year}{2021}.
\newblock \bibinfo{title}{Genomic evidence of a {Sars-Cov-2} reinfection case
  with {E484K} spike mutation in {Brazil}}.
\newblock \bibinfo{journal}{Preprints}
  \DOIprefix\doi{10.20944/preprints202101.0132.v1}.
\bibitem[{Overbaugh(2020)}]{Overbaugh2020}
\bibinfo{author}{Overbaugh, J.}, \bibinfo{year}{2020}.
\newblock \bibinfo{title}{Understanding protection from {SARS-CoV-2} by
  studying reinfection}.
\newblock \bibinfo{journal}{Nature Medicine} \bibinfo{volume}{26},
  \bibinfo{pages}{1680--1681}.
\newblock \DOIprefix\doi{10.1038/s41591-020-1121-z}.
\bibitem[{Perkins and Espa{\~n}a(2020)}]{Perkins2020}
\bibinfo{author}{Perkins, T.A.}, \bibinfo{author}{Espa{\~n}a, G.},
  \bibinfo{year}{2020}.
\newblock \bibinfo{title}{Optimal control of the covid-19 pandemic with
  non-pharmaceutical interventions}.
\newblock \bibinfo{journal}{Bulletin of Mathematical Biology}
  \bibinfo{volume}{82}, \bibinfo{pages}{118}.
\newblock \DOIprefix\doi{10.1007/s11538-020-00795-y}.
\bibitem[{Pontryagin et~al.(1962)Pontryagin, Boltyanskie, Gamkrelidze,
  Mi{\^s}enko and Trirogoff}]{Pontryagin1962}
\bibinfo{author}{Pontryagin, L.}, \bibinfo{author}{Boltyanskie, V.G.},
  \bibinfo{author}{Gamkrelidze, R.V.}, \bibinfo{author}{Mi{\^s}enko, E.F.},
  \bibinfo{author}{Trirogoff, K.N.}, \bibinfo{year}{1962}.
\newblock \bibinfo{title}{The mathematical theory of optimal processes}, in:
  \bibinfo{editor}{Neustadt, L.W.} (Ed.), \bibinfo{booktitle}{Karreman
  Mathematics Research Collection}. \bibinfo{publisher}{Interscience
  Publishers}.
\bibitem[{Rawson et~al.(2020)Rawson, Brewer, Veltcheva, Huntingford and
  Bonsall}]{Rawson2020}
\bibinfo{author}{Rawson, T.}, \bibinfo{author}{Brewer, T.},
  \bibinfo{author}{Veltcheva, D.}, \bibinfo{author}{Huntingford, C.},
  \bibinfo{author}{Bonsall, M.B.}, \bibinfo{year}{2020}.
\newblock \bibinfo{title}{How and when to end the covid-19 lockdown: An
  optimization approach}.
\newblock \bibinfo{journal}{Frontiers in Public Health} \bibinfo{volume}{8},
  \bibinfo{pages}{262}.
\newblock \DOIprefix\doi{10.3389/fpubh.2020.00262}.
\bibitem[{Resende et~al.(2020)Resende, Motta, Roy, Appolinario, Fabri, Xavier,
  Harris, Matos, Caetano, Orgeswalska, Miranda, Garcia, Abreu, Williams, Breuer
  and Siqueira}]{resende2020}
\bibinfo{author}{Resende, P.C.}, \bibinfo{author}{Motta, F.C.},
  \bibinfo{author}{Roy, S.}, \bibinfo{author}{Appolinario, L.},
  \bibinfo{author}{Fabri, A.}, \bibinfo{author}{Xavier, J.},
  \bibinfo{author}{Harris, K.}, \bibinfo{author}{Matos, A.R.},
  \bibinfo{author}{Caetano, B.}, \bibinfo{author}{Orgeswalska, M.},
  \bibinfo{author}{Miranda, M.}, \bibinfo{author}{Garcia, C.},
  \bibinfo{author}{Abreu, A.}, \bibinfo{author}{Williams, R.},
  \bibinfo{author}{Breuer, J.}, \bibinfo{author}{Siqueira, M.M.},
  \bibinfo{year}{2020}.
\newblock \bibinfo{title}{{SARS-CoV-2} genomes recovered by long amplicon
  tiling multiplex approach using nanopore sequencing and applicable to other
  sequencing platforms}.
\newblock \bibinfo{journal}{bioRxiv} \DOIprefix\doi{10.1101/2020.04.30.069039}.
\bibitem[{Rodriguez-Morales et~al.(2020)Rodriguez-Morales, Cardona-Ospina,
  Gutiérrez-Ocampo, Villamizar-Peña, Holguin-Rivera, Escalera-Antezana,
  Alvarado-Arnez, Bonilla-Aldana, Franco-Paredes, Henao-Martinez,
  Paniz-Mondolfi, Lagos-Grisales, Ramírez-Vallejo, Suárez, Zambrano,
  Villamil-Gómez, Balbin-Ramon, Rabaan, Harapan, Dhama, Nishiura, Kataoka,
  Ahmad and Sah}]{RodriguezClinical2020}
\bibinfo{author}{Rodriguez-Morales, A.J.}, \bibinfo{author}{Cardona-Ospina,
  J.A.}, \bibinfo{author}{Gutiérrez-Ocampo, E.},
  \bibinfo{author}{Villamizar-Peña, R.}, \bibinfo{author}{Holguin-Rivera, Y.},
  \bibinfo{author}{Escalera-Antezana, J.P.}, \bibinfo{author}{Alvarado-Arnez,
  L.E.}, \bibinfo{author}{Bonilla-Aldana, D.K.},
  \bibinfo{author}{Franco-Paredes, C.}, \bibinfo{author}{Henao-Martinez, A.F.},
  \bibinfo{author}{Paniz-Mondolfi, A.}, \bibinfo{author}{Lagos-Grisales, G.J.},
  \bibinfo{author}{Ramírez-Vallejo, E.}, \bibinfo{author}{Suárez, J.A.},
  \bibinfo{author}{Zambrano, L.I.}, \bibinfo{author}{Villamil-Gómez, W.E.},
  \bibinfo{author}{Balbin-Ramon, G.J.}, \bibinfo{author}{Rabaan, A.A.},
  \bibinfo{author}{Harapan, H.}, \bibinfo{author}{Dhama, K.},
  \bibinfo{author}{Nishiura, H.}, \bibinfo{author}{Kataoka, H.},
  \bibinfo{author}{Ahmad, T.}, \bibinfo{author}{Sah, R.}, \bibinfo{year}{2020}.
\newblock \bibinfo{title}{Clinical, laboratory and imaging features of
  {COVID-19}: {A} systematic review and meta-analysis}.
\newblock \bibinfo{journal}{Travel Medicine and Infectious Disease} ,
  \bibinfo{pages}{101623}\DOIprefix\doi{10.1016/j.tmaid.2020.101623}.
\bibitem[{Ruktanonchai et~al.(2020)Ruktanonchai, Floyd, Lai, Ruktanonchai,
  Sadilek, Rente-Lourenco, Ben, Carioli, Gwinn, Steele, Prosper, Schneider,
  Oplinger, Eastham and Tatem}]{Ruktanonchaieabc2020}
\bibinfo{author}{Ruktanonchai, N.W.}, \bibinfo{author}{Floyd, J.R.},
  \bibinfo{author}{Lai, S.}, \bibinfo{author}{Ruktanonchai, C.W.},
  \bibinfo{author}{Sadilek, A.}, \bibinfo{author}{Rente-Lourenco, P.},
  \bibinfo{author}{Ben, X.}, \bibinfo{author}{Carioli, A.},
  \bibinfo{author}{Gwinn, J.}, \bibinfo{author}{Steele, J.E.},
  \bibinfo{author}{Prosper, O.}, \bibinfo{author}{Schneider, A.},
  \bibinfo{author}{Oplinger, A.}, \bibinfo{author}{Eastham, P.},
  \bibinfo{author}{Tatem, A.J.}, \bibinfo{year}{2020}.
\newblock \bibinfo{title}{{Assessing the impact of coordinated COVID-19 exit
  strategies across Europe}}.
\newblock \bibinfo{journal}{Science} \DOIprefix\doi{10.1126/science.abc5096}.
\bibitem[{Seow et~al.(2020)Seow, Graham, Merrick, Acors, Pickering, Steel,
  Hemmings, O’Byrne, Kouphou, Galao, Betancor, Wilson, Signell, Winstone,
  Kerridge, Huettner, Jimenez-Guardeño, Lista, Temperton, Snell, Bisnauthsing,
  Moore, Green, Martinez, Stokes, Honey, Izquierdo-Barras, Arbane, Patel, Tan,
  O’Connell, O’Hara, MacMahon, Douthwaite, Nebbia, Batra, Martinez-Nunez,
  Shankar-Hari, Edgeworth, Neil, Malim and Doores}]{Seow2020}
\bibinfo{author}{Seow, J.}, \bibinfo{author}{Graham, C.},
  \bibinfo{author}{Merrick, B.}, \bibinfo{author}{Acors, S.},
  \bibinfo{author}{Pickering, S.}, \bibinfo{author}{Steel, K.J.},
  \bibinfo{author}{Hemmings, O.}, \bibinfo{author}{O’Byrne, A.},
  \bibinfo{author}{Kouphou, N.}, \bibinfo{author}{Galao, R.P.},
  \bibinfo{author}{Betancor, G.}, \bibinfo{author}{Wilson, H.D.},
  \bibinfo{author}{Signell, A.W.}, \bibinfo{author}{Winstone, H.},
  \bibinfo{author}{Kerridge, C.}, \bibinfo{author}{Huettner, I.},
  \bibinfo{author}{Jimenez-Guardeño, J.M.}, \bibinfo{author}{Lista, M.J.},
  \bibinfo{author}{Temperton, N.}, \bibinfo{author}{Snell, L.B.},
  \bibinfo{author}{Bisnauthsing, K.}, \bibinfo{author}{Moore, A.},
  \bibinfo{author}{Green, A.}, \bibinfo{author}{Martinez, L.},
  \bibinfo{author}{Stokes, B.}, \bibinfo{author}{Honey, J.},
  \bibinfo{author}{Izquierdo-Barras, A.}, \bibinfo{author}{Arbane, G.},
  \bibinfo{author}{Patel, A.}, \bibinfo{author}{Tan, M.K.I.},
  \bibinfo{author}{O’Connell, L.}, \bibinfo{author}{O’Hara, G.},
  \bibinfo{author}{MacMahon, E.}, \bibinfo{author}{Douthwaite, S.},
  \bibinfo{author}{Nebbia, G.}, \bibinfo{author}{Batra, R.},
  \bibinfo{author}{Martinez-Nunez, R.}, \bibinfo{author}{Shankar-Hari, M.},
  \bibinfo{author}{Edgeworth, J.D.}, \bibinfo{author}{Neil, S.J.D.},
  \bibinfo{author}{Malim, M.H.}, \bibinfo{author}{Doores, K.J.},
  \bibinfo{year}{2020}.
\newblock \bibinfo{title}{Longitudinal observation and decline of neutralizing
  antibody responses in the three months following {SARS-CoV-2} infection in
  humans}.
\newblock \bibinfo{journal}{Nature Microbiology} \bibinfo{volume}{5},
  \bibinfo{pages}{1598--1607}.
\newblock \DOIprefix\doi{10.1038/s41564-020-00813-8}.
\bibitem[{Tarrataca et~al.(2021)Tarrataca, Dias, Haddad and
  Arruda}]{Tarrataca2021}
\bibinfo{author}{Tarrataca, L.}, \bibinfo{author}{Dias, C.M.},
  \bibinfo{author}{Haddad, D.}, \bibinfo{author}{Arruda, E.F.},
  \bibinfo{year}{2021}.
\newblock \bibinfo{title}{{Flattening the curves: on-off lock-down strategies
  for COVID-19 with an application to Brazil}}.
\newblock \bibinfo{journal}{Journal of Mathematics in Industry}
  \bibinfo{volume}{11}, \bibinfo{pages}{2}.
\newblock \DOIprefix\doi{10.1186/s13362-020-00098-w}.
\bibitem[{Tegally et~al.(2020)Tegally, Wilkinson, Giovanetti, Iranzadeh,
  Fonseca, Giandhari, Doolabh, Pillay, San, Msomi, Mlisana, {von Gottberg},
  Walaza, Allam, Ismail, Mohale, Glass, Engelbrecht, {van Zyl}, Preiser,
  Petruccione, Sigal, Hardie, Marais, Hsiao, Korsman, Davies, Tyers, Mudau,
  York, Maslo, Goedhals, Abrahams, Laguda-Akingba, Alisoltani-Dehkordi, Godzik,
  Wibme, Sewell, Louren{\c c}o, Alcantara, Pond, Weaver, Martin, Lessells,
  Bhiman, Williamson and Oliveira}]{Tegally2020}
\bibinfo{author}{Tegally, H.}, \bibinfo{author}{Wilkinson, E.},
  \bibinfo{author}{Giovanetti, M.}, \bibinfo{author}{Iranzadeh, A.},
  \bibinfo{author}{Fonseca, V.}, \bibinfo{author}{Giandhari, J.},
  \bibinfo{author}{Doolabh, D.}, \bibinfo{author}{Pillay, S.},
  \bibinfo{author}{San, E.J.}, \bibinfo{author}{Msomi, N.},
  \bibinfo{author}{Mlisana, K.}, \bibinfo{author}{{von Gottberg}, A.},
  \bibinfo{author}{Walaza, S.}, \bibinfo{author}{Allam, M.},
  \bibinfo{author}{Ismail, A.}, \bibinfo{author}{Mohale, T.},
  \bibinfo{author}{Glass, A.J.}, \bibinfo{author}{Engelbrecht, S.},
  \bibinfo{author}{{van Zyl}, G.}, \bibinfo{author}{Preiser, W.},
  \bibinfo{author}{Petruccione, F.}, \bibinfo{author}{Sigal, A.},
  \bibinfo{author}{Hardie, D.}, \bibinfo{author}{Marais, G.},
  \bibinfo{author}{Hsiao, M.}, \bibinfo{author}{Korsman, S.},
  \bibinfo{author}{Davies, M.}, \bibinfo{author}{Tyers, L.},
  \bibinfo{author}{Mudau, I.}, \bibinfo{author}{York, D.},
  \bibinfo{author}{Maslo, C.}, \bibinfo{author}{Goedhals, D.},
  \bibinfo{author}{Abrahams, S.}, \bibinfo{author}{Laguda-Akingba, O.},
  \bibinfo{author}{Alisoltani-Dehkordi, A.}, \bibinfo{author}{Godzik, A.},
  \bibinfo{author}{Wibme, C.K.}, \bibinfo{author}{Sewell, B.T.},
  \bibinfo{author}{Louren{\c c}o, J.}, \bibinfo{author}{Alcantara, L.C.J.},
  \bibinfo{author}{Pond, S.L.K.}, \bibinfo{author}{Weaver, S.},
  \bibinfo{author}{Martin, D.}, \bibinfo{author}{Lessells, R.J.},
  \bibinfo{author}{Bhiman, J.N.}, \bibinfo{author}{Williamson, C.},
  \bibinfo{author}{Oliveira, T.}, \bibinfo{year}{2020}.
\newblock \bibinfo{title}{Emergence and rapid spread of a new severe acute
  respiratory syndrome-related coronavirus 2 {(SARS-CoV-2)} lineage with
  multiple spike mutations in {South Africa}}.
\newblock \bibinfo{journal}{medRxiv}
  \DOIprefix\doi{10.1101/2020.12.21.20248640}.
\bibitem[{Tillett et~al.(2021)Tillett, Sevinsky, Hartley, Kerwin, Crawford,
  Gorzalski, Laverdure, Verma, Rossetto, Jackson, Farrell, Van~Hooser and
  Pandori}]{Tillett_2020}
\bibinfo{author}{Tillett, R.L.}, \bibinfo{author}{Sevinsky, J.R.},
  \bibinfo{author}{Hartley, P.D.}, \bibinfo{author}{Kerwin, H.},
  \bibinfo{author}{Crawford, N.}, \bibinfo{author}{Gorzalski, A.},
  \bibinfo{author}{Laverdure, C.}, \bibinfo{author}{Verma, S.C.},
  \bibinfo{author}{Rossetto, C.C.}, \bibinfo{author}{Jackson, D.},
  \bibinfo{author}{Farrell, M.J.}, \bibinfo{author}{Van~Hooser, S.},
  \bibinfo{author}{Pandori, M.}, \bibinfo{year}{2021}.
\newblock \bibinfo{title}{Genomic evidence for reinfection with {SARS-CoV-2}: a
  case study}.
\newblock \bibinfo{journal}{The Lancet Infectious Diseases}
  \bibinfo{volume}{21}, \bibinfo{pages}{52--58}.
\newblock \DOIprefix\doi{10.1016/S1473-3099(20)30764-7}.
\bibitem[{To et~al.(2020)To, Hung, Ip, Chu, Chan, Tam, Fong, Yuan, Tsoi, Ng,
  Lee, Wan, Tso, To, Tsang, Chan, Huang, Kok, Cheng and Yuen}]{to_2020}
\bibinfo{author}{To, K.K.W.}, \bibinfo{author}{Hung, I.F.N.},
  \bibinfo{author}{Ip, J.D.}, \bibinfo{author}{Chu, A.W.H.},
  \bibinfo{author}{Chan, W.M.}, \bibinfo{author}{Tam, A.R.},
  \bibinfo{author}{Fong, C.H.Y.}, \bibinfo{author}{Yuan, S.},
  \bibinfo{author}{Tsoi, H.W.}, \bibinfo{author}{Ng, A.C.K.},
  \bibinfo{author}{Lee, L.L.Y.}, \bibinfo{author}{Wan, P.},
  \bibinfo{author}{Tso, E.Y.K.}, \bibinfo{author}{To, W.K.},
  \bibinfo{author}{Tsang, D.N.C.}, \bibinfo{author}{Chan, K.H.},
  \bibinfo{author}{Huang, J.D.}, \bibinfo{author}{Kok, K.H.},
  \bibinfo{author}{Cheng, V.C.C.}, \bibinfo{author}{Yuen, K.Y.},
  \bibinfo{year}{2020}.
\newblock \bibinfo{title}{{Coronavirus Disease 2019 {(COVID-19)} Re-infection
  by a Phylogenetically Distinct Severe Acute Respiratory Syndrome Coronavirus
  2 Strain Confirmed by Whole Genome Sequencing}}.
\newblock \bibinfo{journal}{Clinical Infectious Diseases}
  \DOIprefix\doi{10.1093/cid/ciaa1275}. \bibinfo{note}{ciaa1275}.
\bibitem[{Vieira et~al.(2021)Vieira, Silva, Garcia, Miranda, Matos, Caetano,
  Resende, Mota, Siqueira and Barth}]{vieira2020}
\bibinfo{author}{Vieira, D.F.B.}, \bibinfo{author}{Silva, M.A.N.},
  \bibinfo{author}{Garcia, C.C.}, \bibinfo{author}{Miranda, M.D.},
  \bibinfo{author}{Matos, A.R.}, \bibinfo{author}{Caetano, B.},
  \bibinfo{author}{Resende, P.C.}, \bibinfo{author}{Mota, F.},
  \bibinfo{author}{Siqueira, M.M.}, \bibinfo{author}{Barth, O.M.},
  \bibinfo{year}{2021}.
\newblock \bibinfo{title}{Morphology and morphogenesis of {SARS-CoV-2 in
  Vero-E6} cells}.
\newblock \bibinfo{journal}{Research Square}
  \DOIprefix\doi{10.21203/rs.3.rs-40432/v1}.
\bibitem[{Voloch et~al.(2020)Voloch, {da Silva}, {de Almeida}, Cardoso,
  Brustolini, Gerber, Guimar{\~a}es, Mariani, Costa, Ferreira, Cavalcanti,
  Frauches, Mello, Galliez, Faffe, Casti{\~n}eiras, Tanuri and
  Vasconcelos}]{Voloch2020}
\bibinfo{author}{Voloch, C.M.}, \bibinfo{author}{{da Silva}, R.F.},
  \bibinfo{author}{{de Almeida}, L.G.}, \bibinfo{author}{Cardoso, C.C.},
  \bibinfo{author}{Brustolini, O.J.}, \bibinfo{author}{Gerber, A.L.},
  \bibinfo{author}{Guimar{\~a}es, A.P.C.}, \bibinfo{author}{Mariani, D.},
  \bibinfo{author}{Costa, R.M.}, \bibinfo{author}{Ferreira, O.C.},
  \bibinfo{author}{Cavalcanti, A.C.}, \bibinfo{author}{Frauches, T.S.},
  \bibinfo{author}{Mello, C.M.B.}, \bibinfo{author}{Galliez, R.M.},
  \bibinfo{author}{Faffe, D.S.}, \bibinfo{author}{Casti{\~n}eiras, T.M.P.},
  \bibinfo{author}{Tanuri, A.}, \bibinfo{author}{Vasconcelos, A.T.R.},
  \bibinfo{year}{2020}.
\newblock \bibinfo{title}{Genomic characterization of a novel {SARS-CoV-2}
  lineage from {Rio de Janeiro, Brazil}}.
\newblock \bibinfo{journal}{medRxiv}
  \DOIprefix\doi{10.1101/2020.12.23.20248598}.

\end{thebibliography}

\scriptsize{

}

\end{document}